\documentclass[aps,prl,twocolumn,superscriptaddress]{revtex4-2}
\usepackage{amssymb}
\usepackage{amsmath}
\usepackage{bm}
\usepackage{color}
\usepackage{float}
\usepackage{graphicx}
\usepackage{hyperref} 

\begin{document}

\title{
Taming Quantum Noise for Efficient Low Temperature Simulations \\ of Open Quantum Systems}
\author{Meng Xu}
\affiliation{
Institute  for Complex Quantum Systems and IQST, Ulm University - Albert-Einstein-Allee 11, D-89069  Ulm, Germany}
\author{Yaming Yan}
\affiliation{Beijing National Laboratory for Molecular Sciences, State Key Laboratory for Structural Chemistry of Unstable
and Stable Species, Institute of Chemistry, Chinese Academy of Sciences, Zhongguancun, Beijing 100190, China and
University of Chinese Academy of Sciences, Beijing 100049, China}
\author{Qiang Shi}
\affiliation{Beijing National Laboratory for Molecular Sciences, State Key Laboratory for Structural Chemistry of Unstable
and Stable Species, Institute of Chemistry, Chinese Academy of Sciences, Zhongguancun, Beijing 100190, China and
University of Chinese Academy of Sciences, Beijing 100049, China}
\author{J. Ankerhold}
\affiliation{
Institute  for Complex Quantum Systems and IQST, Ulm University - Albert-Einstein-Allee 11, D-89069  Ulm, Germany}
\author{J. T. Stockburger}
\affiliation{
Institute  for Complex Quantum Systems and IQST, Ulm University - Albert-Einstein-Allee 11, D-89069  Ulm, Germany}

\date{\today}
\begin{abstract}
The hierarchical equations of motion (HEOM), derived from the exact Feynman-Vernon path integral, is one of the most powerful numerical methods to simulate the dynamics of open quantum systems. Its applicability has so far been limited to specific forms of spectral reservoir distributions and relatively elevated temperatures. Here we solve this problem and introduce an effective treatment of quantum noise in frequency space by systematically clustering higher order Matsubara poles, equivalent to an optimized rational decomposition. This leads to an elegant extension of the  HEOM to arbitrary temperatures and very general reservoirs in combination with efficiency, high accuracy, and long-time stability. Moreover, the technique can directly be implemented in other approaches such as Green's function, stochastic, and pseudomode formulations. As one highly nontrivial application,  for the subohmic spin-boson model at vanishing temperature the Shiba relation is quantitatively verified which predicts the long-time decay of correlation functions.
\end{abstract}

\maketitle
\newpage

\emph{Introduction.-} Deviations from ideal unitary evolution of finite quantum systems are ubiquitous \cite{gardiner2004quantum,breuer02,weiss12}---the exchange of information \cite{nielsen2010quantum,averin2001macroscopic}, energy \cite{nitzan06,may11}, and particles \cite{datta95} with environments are never completely suppressed. The effects caused by this exchange are often marginal or incidental, as in the broadening of spectral lines. In other cases, they are dominant elements of a physical effect, e.g., charge transport in polarizable media, or even a key element of function, providing the required directivity of charge or exciton transfer in solid state \cite{van1991quantum} and  biological systems \cite{mohseni2014quantum}. In quantum information processing, on the other hand, environmental effects have a marked detrimental effect on performance due to decoherence effects. Open quantum systems with significant system-environment correlations also figure prominently in the field of quantum phase transitions \cite{bulla2003numerical,winter2009quantum}; the archetypal spin-boson system was one of the first for which the famed quantum-classical mapping was performed~\cite{spohn89}. 
While formally exact methods are needed to describe these systems, other settings with moderate system-reservoir correlations can be accommodated using master equations beyond the Born-Markov approximation, e.g., based on the small-polaron transformation \cite{jang08,jang09} or modified Lindblad equations \cite{palmieri09,abramavicius2010quantum}.

When open quantum systems are probed by complex driving patterns, such as in multidimensional spectroscopy \cite{mukamel_book}, few methods provide a reliable theoretical picture of the underlying reduced-state dynamics. These include numerical path integral simulations \cite{muhlbacher2004path, thorwart09} and stochastic representations of reservoir fluctuations and response~\cite{stockburger02,stockburger04,hartmann2017exact,stock16a}. 
In addition, dynamical modeling based on hierarchical equations of motion (HEOM) has emerged as one of the most popular and reliable methods \cite{tanimura89,tanimura1993two,tanimura06,abramavicius2009coherent,ishizaki09b,sarovar2012reduced,yan2014theory,tanimura2020numerically}. While it retains the structural simplicity of a linear dynamical system, its state space, augmented by auxiliary density operators, is complex enough to treat effects which delicately depend on system-reservoir correlations. However, this success story has yet to be extended from chemical and biological systems at elevated temperature to equally challenging applications in the growing field of quantum technologies. This includes quantum information processing, quantum thermodynamics, and quantum sensing, to name but a few, operating at very low temperatures, in the presence of structured environments \cite{remark-struc}, and on longer timescales.

As in many other approaches \cite{makri1995tensor,stockburger02,shao2004decoupling,cohen2005taming,kast2013persistence,suess2014hierarchy,tamascelli2018nonperturbative,cygorek2017nonlinear,strathearn2018efficient,pollock2018non,bose2022multisite,cygorek2022simulation} to open quantum systems, the reservoir correlation function $C(t)$ characterizing the effect of both fluctuations and dissipation, is a cornerstone of the theory. In HEOM, it must be represented using a finite set of basis functions \cite{xu05,duan17,rahman2019chebyshev,cui2019highly,ikeda2020generalization,chen2022universal}. In the most common approach, the basis functions are decaying exponentials, which result, essentially, from Wick-rotating oscillatory Matsubara-frequency terms of thermal Green's functions. The low-temperature limit of this procedure leads to an exponential proliferation of these terms making the HEOM too resource intensive for any practical purpose, even when extended by tensor network representations \cite{shi2018efficient,borrelli2019density,yan2021efficient}. Physically, this problem is due to scale-free noise spectra at low frequencies associated with zero-point quantum fluctuations.  Complications experienced by other approaches can be traced back to this problem as well, e.g., in Green's function techniques \cite{croy09,beach2000reliable,gu2020generalized}, the hierarchy of stochastic pure state (HOPS) method
\cite{suess2014hierarchy}, and the pseudomode formulation in extended Hilbert space \cite{tamascelli2018nonperturbative,pleasance2020generalized,trivedi2021convergence}. 

Here we overcome this fundamental difficulty by concentrating entirely on the real-time function $C(t)$ and its spectrum on the real frequency axis. In the spirit of hidden Markov models~\cite{Awad2015}, we treat any auxiliary quantities not directly observable---including the analytic pole structure of the reservoir noise power spectrum---as adjustable within tight constraints of accuracy on the observable quantities, e.g., the reduced density matrix as well as some system-reservoir correlations. Since numerical analytic continuation is an ill-posed problem, we have considerable freedom to select an optimized pole structure quite different from Matsubara poles while maintaining high accuracy down to zero temperature. Where the required computational resources previously diverged exponentially with inverse temperature $\beta$, we here offer an efficient platform to explore open-system dynamics in fine detail down to $T=0$ and for  very general spectral reservoir distributions. We suggest the name free-pole-HEOM (FP-HEOM) for this dynamical framework.

As a demonstration of the effectiveness of FP-HEOM, we provide highly accurate simulations for a subohmic spin-boson model, which defies a perturbative treatment due to the large spectral weight of low-frequency modes and has been thoroughly analyzed in the context of quantum phase transitions \cite{weiss12}. To illustrate the stability and accuracy of the method, we determine the long-time algebraic tails of the zero-temperature spin-spin correlation function, where our results perfectly match the Shiba relation \cite{shiba75a,sasse90b} linking it to $C(t)$. Equally important, the optimized pole structure of FP-HEOM will directly benefit other approaches mentioned above, e.g., Green's function and HOPS methods~\cite{suess2014hierarchy}, and pseudomode formulations~\cite{tamascelli2018nonperturbative}. 

\emph{Quantum noise in open system dynamics.-} The common modeling of open quantum systems starts from a system+reservoir Hamiltonian $H=H_S+ Q_S X_R+ H_R$, where for the sake of simplicity we assume a bilinear coupling between a system $H_S$ with operator $Q_S$ and a reservoir $H_R$ with $X_R$ \cite{breuer02,weiss12}. The time evolution of the reduced density operator with a factorized initial state is then given by
\begin{equation}\label{eq:density}
    \rho(t) = {\rm Tr}_R\{{\rm e}^{-iHt}\rho(0)\otimes W_\beta\,  {\rm e}^{iHt}] \;\;,
\end{equation}
with the equilibrium density of the bare reservoir $W_\beta=\exp(-\beta H_R)/{\rm Tr}\{\exp(-\beta H_R\}$ at inverse temperature $\beta=1/T$ and $\hbar=k_{\rm B} = 1$.

\begin{figure}[htbp]
\centering
\includegraphics[width=8.6cm]{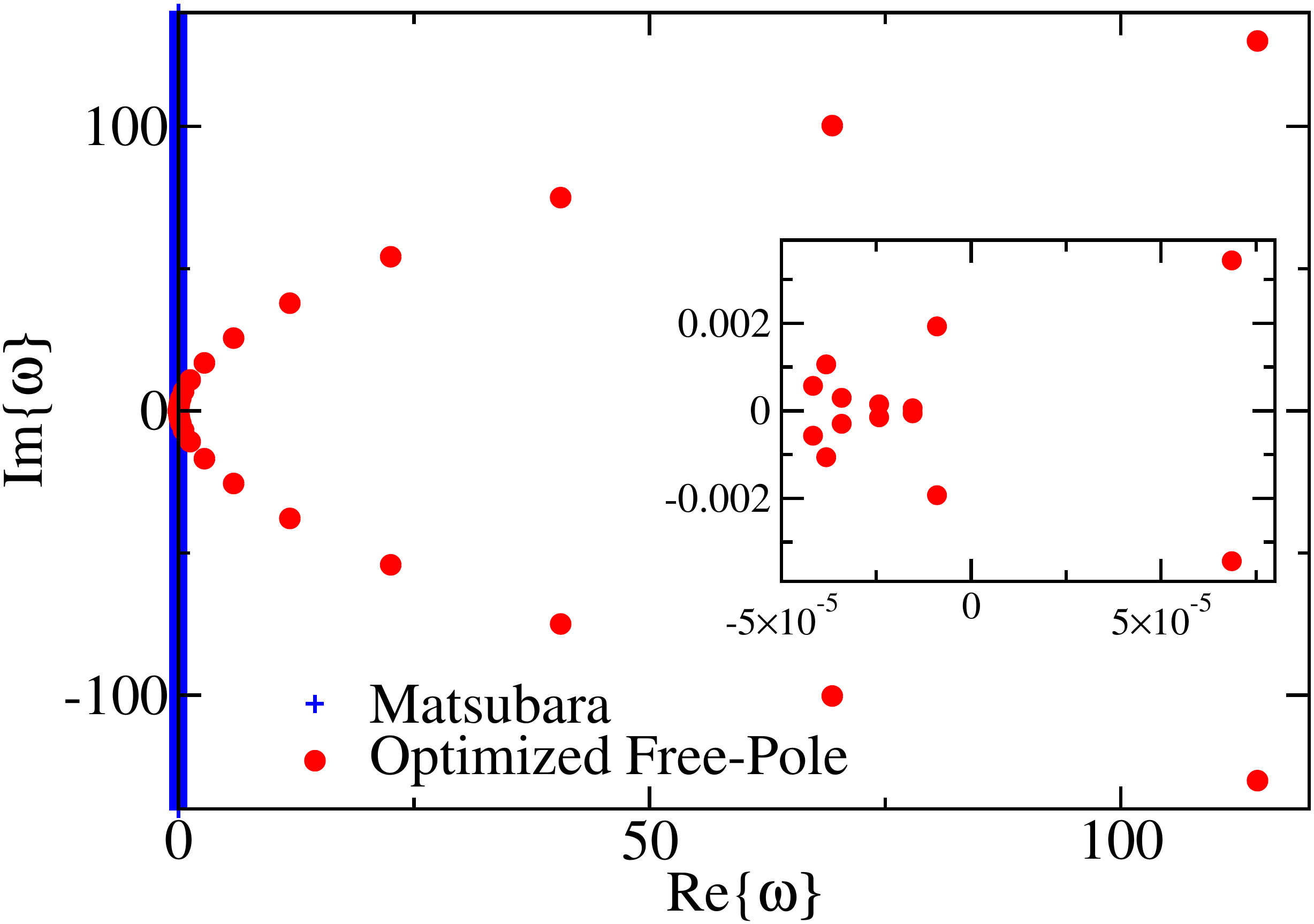}
\caption{Matsubara poles of the spectral noise power $S_\beta(\omega)$ of a thermal reservoir at $T=0$, where they merge into a branch cut along the imaginary axis (blue) together with the optimized pole distribution (red) based on its rational decomposition [Eq. (\ref{Eq:rationbary})]. For the latter, data refer to a subohmic spectral density [Eq. (\ref{eq:subohmic})] with $s=1/2$, $\alpha=0.05$, and $\omega_c=20$ (in arbitrary units) with an accuracy $\delta= 10^{-8}$ in the domain $\mathcal{A}=[-10^{3},-10^{-4}] \cup [10^{-4},10^{3}]$ \cite{SI}. Inset: low frequency range. }
\label{fig1}
\end{figure}

Reservoirs with a macroscopic number of degrees of freedom are characterized by Gaussian fluctuations, i.e.\ by the autocorrelation function $C(t)=\langle X_R(t) X_R(0)\rangle_R$ given that $\langle X_R\rangle_R=0$ and $\langle \cdot\rangle_R={\rm Tr}\{W_\beta\, \cdot\}$. Here we focus on the case of a bosonic thermal reservoir, characterized by a noise spectrum
\begin{equation}\label{Eq:fct}
    S_{\beta}(\omega) = \int_{-\infty}^{+\infty}dt \, {\rm e}^{i\omega t} C(t),
\end{equation}
where $S_{\beta}(\omega)$ and $S_{\beta}(-\omega)$ are related by the fluctuation-dissipation theorem which can be represented as $S_{\beta}(\omega) = 2 [n_\beta(\omega) +1]J(\omega)$. Here, $n_\beta(\omega)=1/[\exp(\beta\omega)-1]$ is the Bose distribution, and the spectral density $J(\omega)$ is an antisymmetric function with finite bandwidth characterized by a cut-off frequency $\omega_c$.
Most previous work focused on analytic functions $J(\omega)$, leading to a canonical pole structure of $S_{\beta}(\omega)$ from which one arrives at  $C(t)=C_J(t)+C_\beta(t)$, where  $C_J$ collects the contributions from poles of $J(\omega)$ while $C_\beta(t)=\sum_{n= 1}^\infty c_n {\rm e}^{-\nu_n t}$ carries the poles of $n_\beta(\omega)$ along the imaginary axis, i.e., the Matsubara frequencies  $\nu_n=2\pi n/\beta$~\footnote{The canonical structure has no pole at $\omega=0$.}. In the high-temperature limit, $C_\beta(t)$ drops exponentially fast compared with $C_J(t)$ and can thus be neglected. However, with decreasing temperature an increasing number of Matsubara frequencies is relevant, and $C(t)$ tends to be dominated by quantum noise.  In fact, for $T\to 0$, an infinite number of Matsubara frequencies contribute (see Fig.~\ref{fig1}) and lead to the well-known algebraic long-time tails in $C(t)$ \cite{weiss12}. As a consequence, time retardation induces long-range self-interactions in the reduced dynamics [Eq. (\ref{eq:density})] and renders any simple time-local equation of motion for $\rho(t)$ impossible.

To account for quantum noise in practical simulations, conventional HEOM as well as some other nonperturbative approaches require a decomposition of $C(t)$ into {\it finite} number of exponential terms, implying an upper limit $K$ on the Matsubara index $k$. This, in turn, restricts the accuracy of the decomposition to temperatures high enough such that $\nu_K t\gg 1$ for all relevant timescales. This way, in HEOM the time retardation is "unraveled" by introducing auxiliary density operators (ADOs) $\rho_{n_1,\ldots,n_K}(t)$ with $\rho(t)\equiv \rho_{0,\ldots,0}(t)$ to construct a nested hierarchy of time-local equations of motion \cite{tanimura06},
\begin{equation}\label{eq:heom}
    \dot{\rho}_{\bf n}(t)= -i \mathcal{L}_S \rho_{\bf n}(t) + \sum_{\bf {n'}}\mathcal{D}_{\bf n,n'}\, \rho_{\bf n'}(t)\, .
\end{equation}
Here $\mathcal{L}_S$ is the system Liouvillian, and the entries of the tensor $\mathcal{D}$, representing all dissipative effects, are either scalar damping terms or linear functions of commutators and anticommutators involving $Q_S$. The subscript $\bm{n} = \{n_1,\ldots,n_k,\ldots n_K\}$ indicates the $n_k$th hierarchy order in $C(t)$ for the $k$th Matsubara frequency.

The canonical form of HEOM does not apply to general mode distributions $J(\omega)$ and suffers from the even more severe restriction to the regime of moderate to high temperatures, due to the excessive growth of required computational resources with $K$ \cite{shi2009efficient}. Recent extensions partially address this problem by using different types of memory basis functions \cite{xu05,hu11,duan17,erpenbeck2018extending,rahman2019chebyshev,cui2019highly,ikeda2020generalization,chen2022universal}. Yet, the problem remains and has been the main obstacle to efficiently explore open quantum dynamics for general reservoirs in the low temperature domain and for long times. 

\emph{Free-pole HEOM.-}
One key to finding a highly accurate yet minimal pole structure for $S_\beta(\omega)$ is the barycentric representation of rational functions~\cite{nakatsukasa2018aaa}
\begin{equation}\label{Eq:rationbary}
 \tilde{S}_{\beta} (\omega) = \left. \sum_{j}^{m} 
 \frac{W_j S_{\beta}(\Omega_j)}{\omega-\Omega_j}   \middle/
 \sum_{j}^{m}\frac{W_j}{\omega - \Omega_j} \right.
\end{equation}
with $\tilde S_\beta(\Omega_j) = S_\beta(\Omega_j)$. The key steps of the pole optimization algorithm are as follows: A very large set of potential support points $\Omega_j$ is chosen, which is dense enough to represent the entire real axis. The optimized pole structure is created iteratively from $m=1$ with two steps in each iteration~\cite{nakatsukasa2018aaa,nakatsukasa2020algorithm}. Step 1 consists of finding $W_j$ which minimize the mean square error of $\tilde S_\beta(\omega)$ on the set of \emph{all} $\Omega_j$. Using these weights, step 2 adds the $\Omega_j$ with largest error to the set used in Eq.~(\ref{Eq:rationbary}). If it is not needed to achieve the required tolerance, the algorithm terminates. Even at $T=0$, the number of poles at which the procedure terminates grows rather slowly with increasing demand for accuracy; for details see Fig. 2 in the Supplemental Material \cite{SI}. With pole locations and residues determined by the algorithm, the multiexponential form $(t\geq 0)$
\begin{equation}
\label{Eq:cbary}
    C(t) = \sum_{k=1}^{K} \,d_k\,  {\rm e}^{-i\,\omega_k t-\gamma_k t}
\end{equation}
is recovered from poles of $\tilde S_\beta(\omega)$ located in the lower complex half plane. Further details of the algorithm can be found in the Supplemental Material \cite{SI}. An example is shown in Fig.~\ref{fig1} for a subohmic model $J(\omega)\propto \omega^{1/2}$ at $T=0$ with $K=31$ and high accuracy $\delta=10^{-8}$ \cite{SI}. The distribution of quasimodes reflects the crucial low-frequency portion of $J(\omega)$, while it becomes sparse at higher frequencies with stronger "damping" (larger $\gamma_k$) reflecting a finite cut-off frequency.

Based on the representation [Eq. (\ref{Eq:cbary})], we now start from the
Feynman-Vernon path integral representation of (\ref{eq:density}) to derive  the new FP-HEOM, adapted to complex exponentials,
\begin{equation}\label{Eq:fpheom}
\begin{split}
   \dot{\hat{\rho}}_{\bf m,n} =& -i\mathcal{L}_S\hat{\rho}_{\bf m,n} -\sum_{k=1}^K (m_k z_k+n_k z_k^*) \hat{\rho}_{\bf m,n} \\
      & -i\sum_{k=1}^K \mathcal{L}_k^{+} \hat{\rho}_{\bm{m,n}} -i\sum_{k=1}^K \mathcal{L}_k^{-} \hat{\rho}_{\bm{m,n}} \;\;
\end{split}
\end{equation}
with multi-index $(\bm{m,n}) \equiv \{m_1,\ldots,m_K,n_1,\ldots,n_K\}$ associated with forward and backward system path, complex-valued coefficients $z_k \equiv \gamma_k + i\omega_k$, 
and raising and lowering superoperators $\mathcal{L}_k^+$ and $\mathcal{L}_k^-$, acting on the $k$th quasimode in Eq. (\ref{Eq:cbary}), again involving $Q_S$~\cite{SI}.
This FP-HEOM  generalizes known approaches and reduces to them in limiting cases, namely,
to conventional HEOM versions for $\omega_k=0$ \cite{tanimura89} and to the mixed quantum-classical Liouville equation \cite{liu14} for $\gamma_k = 0$. Equation (\ref{Eq:fpheom}) together with Eqs. (\ref{Eq:cbary}) and (\ref{Eq:rationbary}) is the main result of this Letter. It constitutes the FP-HEOM as a breakthrough to efficiently explore open quantum dynamics also at very low temperatures, long times, and for general reservoir mode distributions.

\emph{Subohmic spin boson model.-}
In order to illustrate this, we consider the full relaxation dynamics of the subohmic spin-boson model with total Hamiltonian 
\begin{equation*}\label{Eq:hspm}
    H = \frac{\epsilon}{2}\sigma_z + \Delta\sigma_x + \sum_j c_j x_j\sigma_z  + \sum_j\left(\frac{p_j^2}{2m_j} + \frac{1}{2}m_j\omega_j^2 x_j^2 \right)\, .
\end{equation*}
Here, $\sigma_{x,z}$ denote Pauli matrices of a general two level system (TLS) with intersite coupling $\Delta$ and detuning $\epsilon$. The TLS interacts with a reservoir of harmonic modes $p_j, x_j$ with subohmic distribution $J(\omega) = (\pi/2)\sum_j(c_j^2/m_j\omega_j)\delta(\omega-\omega_j)$ which in the continuum limit reads as
\begin{equation}\label{eq:subohmic}
    J(\omega) = \frac{\pi}{2}\,\alpha\, \omega_c^{1-s}\,\omega^{s}\, {\rm e}^{-\omega/\omega_c}\,, \, s\leq 1\;\;
\end{equation}
with large cutoff $\omega_c$ and dimensionless coupling  $\alpha$.

The interplay of the relatively large portion of low-frequency reservoir modes $J(\omega)\propto \omega^s$  with the internal TLS dynamics makes this model extremely challenging to simulate, particularly in the long time limit and close to or at $T=0$. Indeed, its low temperature properties have received substantial attention as a generic model for dissipation induced quantum phase transitions (see e.g., Refs.~\cite{bulla2003numerical,winter2009quantum}) and for quantum noise properties in solid state systems such as metal rings \cite{tong2006signatures} and superconducting circuits \cite{shnirman2002,yamamoto2019microwave}. While the literature is numerous and substantial, one of the cornerstone predictions of open quantum systems, the Shiba relations \cite{shiba75a,sasse90b}, have not been numerically reproduced yet due to the lack of proper simulation techniques. They state that for asymptotic times, the two-point correlator  $S_{zz}(t) = \langle \sigma_z(t)\sigma_z(0) + \sigma_z(0)\sigma_z(t)\rangle/2$ is slaved to follow the reservoir correlation $C(t)$, i.e.,
\begin{equation}
\label{eq:shiba}
   \left. S_{zz}(t)\right|_{T=0} = \frac{\bar{\chi}_z^2}{4}\,\Re\, C(t)\;\; 
\end{equation}
with $\bar{\chi}_z=2 (\partial/\partial\epsilon) \langle\sigma_z(t\to \infty)\rangle$. In case of Eq. (\ref{eq:subohmic}),  one finds \cite{weiss12} for $\omega_c t\gg 1$  that
\begin{equation}
\Re\, C(t) = - \xi_s\, \omega_c^{1-s}/t^{1+s},
\end{equation}
thus indicating the long-time algebraic decay due to summation over {\em all} Matsubara frequencies in Eq. (\ref{Eq:fct}), with prefactor $\xi_s=2 \alpha s (1-s)\Gamma(s-1) \cos\left[(1+s)(\pi/2)\right]$.  Explicitly, Eq. (\ref{eq:shiba}) then predicts in the regime $\alpha\ll 1$ that
\begin{equation}\label{eq:shiba-subohmic}
\left. S_{zz}(t)\right|_{T=0}=-\xi_s \left(\frac{\omega_c}{\Delta}\right)^{1-s}\,  \frac{1}{(\Delta t)^{1+s}}\, ,
\end{equation}
which is highly nontrivial to validate numerically. 
\begin{figure}[htbp]
\centering
\includegraphics[width=8.6cm]{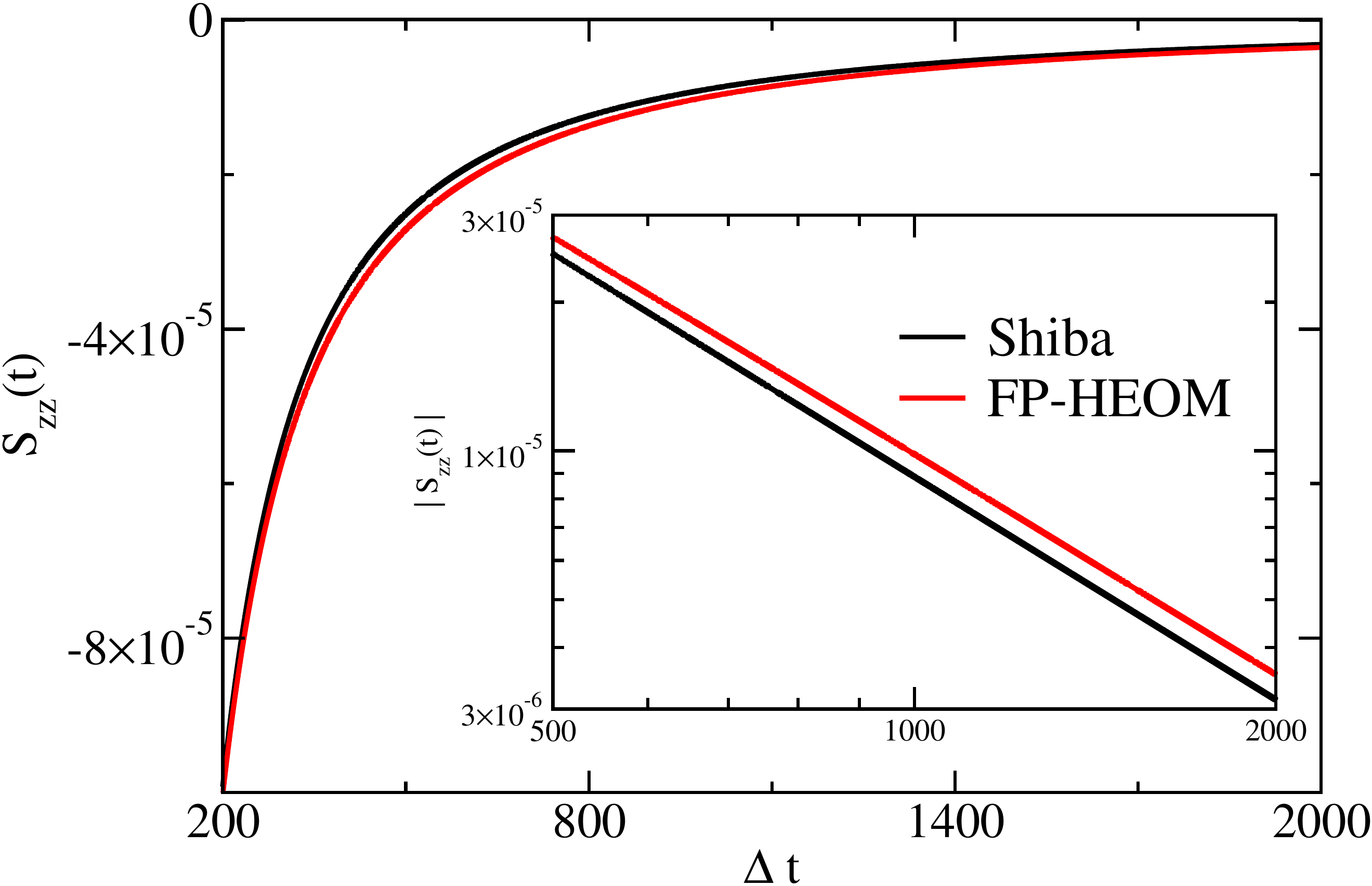}
\caption{Validation of the Shiba relation [Eq. (\ref{eq:shiba-subohmic})] of the subohmic spin boson model [Eq. \ref{eq:subohmic}] with $\epsilon=0$, $\Delta=1$ (in arbitrary units), and  $s=1/2$ in the long time domain. Inset: long time behavior on a logarithmic scale. Other parameters are as in Fig.~\ref{fig1}.} 
\label{fig2}
\end{figure}

As clearly shown in Fig.~\ref{fig2} for $s=1/2$, the FP-HEOM can capture the power law behavior with an accuracy of better than $10^{-5}$ \footnote{The small shift between the analytical prediction and FP-HEOM is mainly due to the $\alpha\to 0$-limit of the prefactor in Eq.(\ref{eq:shiba-subohmic})}. We emphasize both, the asymptotic time range $\Delta t\gg 1$ required for all transients to have died out, and the resolution of the numerics at these small absolute values of $S_{zz}(t)$ starting from $S_{zz}(t=0)=1$. Similar accuracy also applies to other exponents $s$ as shown in the Supplemental Material \cite{SI}. Since computational costs grow exponentially with the number of required poles, the FP-HEOM is a real game changer: compared with at least $10^4$ Matsubara poles for conventional HEOM, the FP-HEOM needs just $K=31$. In the examples shown in this Letter, its performance is further enhanced by a matrix product state representation of the coupled dynamics of the ADOs, resulting in a computational complexity which is only linear in the number of poles $K$ \cite{shi2018efficient}. For further information on the computational complexity, see the Supplemental Material \cite{SI}.

\begin{figure}[htbp]
\centering
\includegraphics[width=8.6cm]{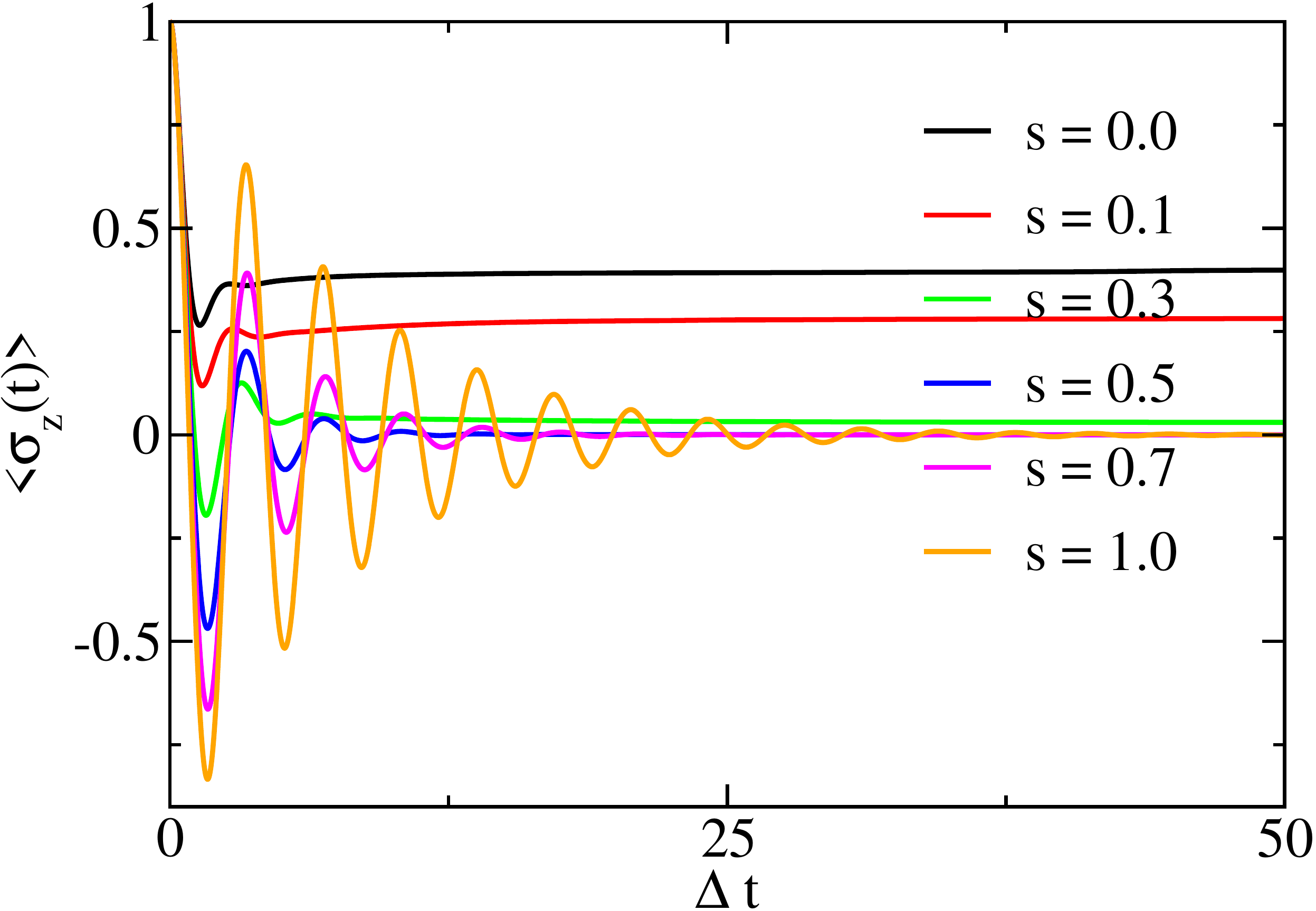}
\caption{Relaxation dynamics of the subohmic spin boson model for various spectral exponents $s$ at $T=0$ displaying the transition from a delocalized ($\langle\sigma_z\rangle=0)$ to a localized ($\langle\sigma_z\rangle\neq 0)$ asymptotic state. Parameters are as in Figs.~\ref{fig1} and \ref{fig2}.}
\label{fig3}
\end{figure}

Equipped with the FP-HEOM, we can now also capture the quantum phase transition from delocalization $\langle\sigma_z(t\to \infty)\rangle=0$ to localization $\langle\sigma_z(t\to \infty)\rangle\neq 0$, for example, at fixed $\alpha$ but varying exponent $s$. Fig.~\ref{fig3} depicts this behavior and focuses on the short to moderately long times to highlight that this substantially advanced method covers the quantum dynamics on all timescales. For $s>0.3$, the TLS delocalizes, cf.\, Fig.~\ref{fig2}, Ref. \cite{SI}, while for values below, the strong portion of low frequency modes destroys coherence via long-range correlations in time to induce reservoir-dressed equilibrium states, as shown, e.g., using imaginary time path integral Monte Carlo simulations \cite{winter2009quantum}.

\begin{figure}[htbp]
\centering
\includegraphics[width=8.6cm]{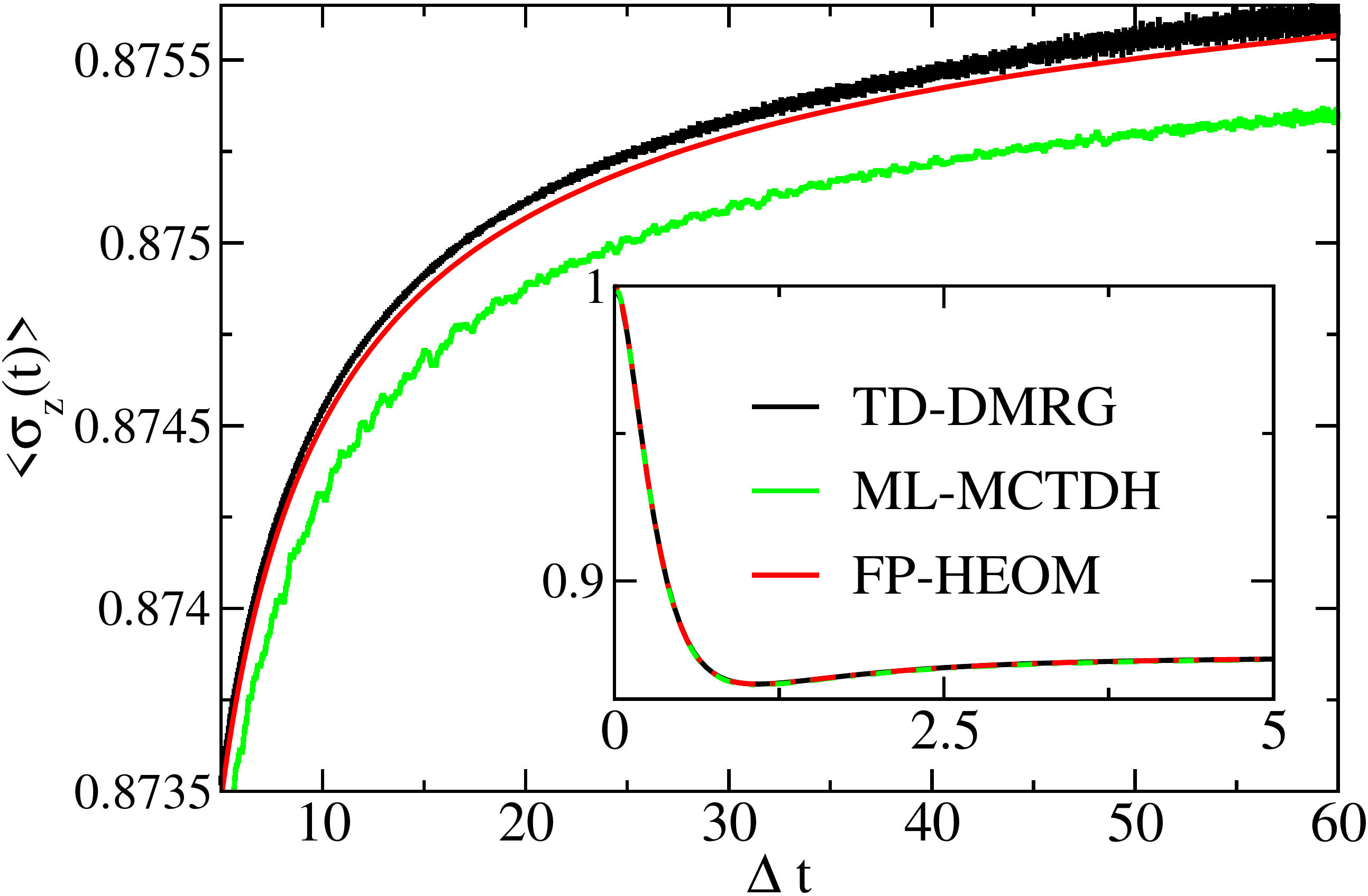}
\caption{Long-time dynamics of the subohmic spin boson model with $s=1/2$ at $T=0$ and strong coupling $\alpha=0.4$ as predicted by TD-DMRG (black, 1000 modes), ML-MCTDH (green, 1000 modes) and FP-HEOM (red, 21 FP-modes; accuracy corresponds to line width). On short time scales, all approaches coincide (inset). Parameters are as in Figs.~\ref{fig1} and \ref{fig2}.}
\label{fig4}
\end{figure}
We briefly compare in Fig.~\ref{fig4} FP-HEOM reliability with other simulation techniques, concentrating on a subohmic case in a highly challenging regime, namely, strong coupling at $T=0$ with dominant memory effects, long propagation times, and high requirements. While reduced-state dynamics, e.g., based on path integral methods, can effectively deal with the first criteria, they suffer from a prohibitive performance degradation when the latter two are emphasized. With no analytic predictions available, numerical data, however, have been previously obtained by two advanced basis set methods formulated in full Hilbert space, i.e.\ the multilayer multiconfiguration time-dependent Hartree (ML-MCTDH) method \cite{wang10from} and the time-dependent density matrix renormalization group (TD-DMRG) \cite{ren2022time}.  Apparently, their results for the long-time dynamics seems more noisy than the FP-HEOM, mostly due to the need for a more pronounced basis set truncation. Our new HEOM propagation, based on only a few tens of quasimodes, appears numerically very stable throughout parameter space and at substantially reduced numerical costs, for details see the Supplemental Material \cite{SI}. More generally, it seems that the FP-HEOM based on the barycentric representation represents an "optimized unraveling" in an extended space when only the reduced dynamics is of interest.

\emph{Conclusion.-}In this Letter we have presented the FP-HEOM as a new tool to efficiently explore the most challenging regimes for the dynamics of open quantum systems derived from the exact Feynman-Vernon path integral. It opens new avenues for the HEOM approach in that it can now also be used to explore the regimes of low and ultralow temperatures resolving even minute effects in asymptotic time domains, in addition to their proven strength in the case of structured reservoirs \cite{kreisbeck2012long,shi2018efficient}. 
The essential ingredient is an effective clustering of the closely spaced Matsubara modes via an optimized rational decomposition of the spectral noise power.  
This leads to a representation of the time-dependent reservoir correlation in a moderate number of quasimodes even at $T=0$ and without restrictions to a time window. Since the FP-HEOM combines high accuracy with high computational efficiency, it provides the basis for high-precision simulations in reasonable computational time required for current quantum technological devices.
Moreover, the pole-clustering methodology can easily be implemented into other approaches to boost their performance such as the Green's function \cite{croy09,beach2000reliable,gu2020generalized}, the HOPS \cite{suess2014hierarchy}, and the non-perturbative pseudomodes \cite{tamascelli2018nonperturbative,pleasance2020generalized}. 

\emph{Acknowledgements.-} We acknowledge fruitful discussions with Frithjof B. Anders, Matthias Vojta, Haobin Wang, Jiajun Ren, and Houdao Zhang and thank them for sharing data with us. M. X. acknowledges support from the state of Baden-Württemberg through bwHPC (JUSTUS 2). This work has been supported by the IQST, the German Science Foundation (DFG) under AN336/12-1 (For2724), the State of Baden-W{\"u}ttemberg under KQCBW/SiQuRe and the BMBF through QSolid. Y. Y. and Q. S. acknowledge the support from NSFC (Grant No. 21933011) and the K. C. Wong Education Foundation.
\section*{References}
\bibliography{FPHEOM}

\begin{thebibliography}{80}%
\makeatletter
\providecommand \@ifxundefined [1]{%
 \@ifx{#1\undefined}
}%
\providecommand \@ifnum [1]{%
 \ifnum #1\expandafter \@firstoftwo
 \else \expandafter \@secondoftwo
 \fi
}%
\providecommand \@ifx [1]{%
 \ifx #1\expandafter \@firstoftwo
 \else \expandafter \@secondoftwo
 \fi
}%
\providecommand \natexlab [1]{#1}%
\providecommand \enquote  [1]{``#1''}%
\providecommand \bibnamefont  [1]{#1}%
\providecommand \bibfnamefont [1]{#1}%
\providecommand \citenamefont [1]{#1}%
\providecommand \href@noop [0]{\@secondoftwo}%
\providecommand \href [0]{\begingroup \@sanitize@url \@href}%
\providecommand \@href[1]{\@@startlink{#1}\@@href}%
\providecommand \@@href[1]{\endgroup#1\@@endlink}%
\providecommand \@sanitize@url [0]{\catcode `\\12\catcode `\$12\catcode
  `\&12\catcode `\#12\catcode `\^12\catcode `\_12\catcode `\%12\relax}%
\providecommand \@@startlink[1]{}%
\providecommand \@@endlink[0]{}%
\providecommand \url  [0]{\begingroup\@sanitize@url \@url }%
\providecommand \@url [1]{\endgroup\@href {#1}{\urlprefix }}%
\providecommand \urlprefix  [0]{URL }%
\providecommand \Eprint [0]{\href }%
\providecommand \doibase [0]{https://doi.org/}%
\providecommand \selectlanguage [0]{\@gobble}%
\providecommand \bibinfo  [0]{\@secondoftwo}%
\providecommand \bibfield  [0]{\@secondoftwo}%
\providecommand \translation [1]{[#1]}%
\providecommand \BibitemOpen [0]{}%
\providecommand \bibitemStop [0]{}%
\providecommand \bibitemNoStop [0]{.\EOS\space}%
\providecommand \EOS [0]{\spacefactor3000\relax}%
\providecommand \BibitemShut  [1]{\csname bibitem#1\endcsname}%
\let\auto@bib@innerbib\@empty
\bibitem [{\citenamefont {Gardiner}\ and\ \citenamefont
  {Zoller}(2004)}]{gardiner2004quantum}%
  \BibitemOpen
  \bibfield  {author} {\bibinfo {author} {\bibfnamefont {C.~W.}\ \bibnamefont
  {Gardiner}}\ and\ \bibinfo {author} {\bibfnamefont {P.}~\bibnamefont
  {Zoller}},\ }\href@noop {} {\emph {\bibinfo {title} {Quantum noise: a
  handbook of {M}arkovian and non-{M}arkovian quantum stochastic methods with
  applications to quantum optics}}},\ \bibinfo {edition} {3rd}\ ed.\ (\bibinfo
  {publisher} {Springer},\ \bibinfo {address} {Berlin},\ \bibinfo {year}
  {2004})\BibitemShut {NoStop}%
\bibitem [{\citenamefont {Breuer}\ and\ \citenamefont
  {Petruccione}(2002)}]{breuer02}%
  \BibitemOpen
  \bibfield  {author} {\bibinfo {author} {\bibfnamefont {H.~P.}\ \bibnamefont
  {Breuer}}\ and\ \bibinfo {author} {\bibfnamefont {F.}~\bibnamefont
  {Petruccione}},\ }\href@noop {} {\emph {\bibinfo {title} {The Theory of Open
  Quantum Systems}}}\ (\bibinfo  {publisher} {Oxford University Press},\
  \bibinfo {address} {New York},\ \bibinfo {year} {2002})\BibitemShut {NoStop}%
\bibitem [{\citenamefont {Weiss}(2012)}]{weiss12}%
  \BibitemOpen
  \bibfield  {author} {\bibinfo {author} {\bibfnamefont {U.}~\bibnamefont
  {Weiss}},\ }\href@noop {} {\emph {\bibinfo {title} {Quantum dissipative
  systems}}},\ \bibinfo {edition} {4th}\ ed.\ (\bibinfo  {publisher} {World
  Scientific},\ \bibinfo {address} {New Jersey},\ \bibinfo {year}
  {2012})\BibitemShut {NoStop}%
\bibitem [{\citenamefont {Nielsen}\ and\ \citenamefont
  {Chuang}(2010)}]{nielsen2010quantum}%
  \BibitemOpen
  \bibfield  {author} {\bibinfo {author} {\bibfnamefont {M.~A.}\ \bibnamefont
  {Nielsen}}\ and\ \bibinfo {author} {\bibfnamefont {I.~L.}\ \bibnamefont
  {Chuang}},\ }\href {https://doi.org/10.1017/CBO9780511976667} {\emph
  {\bibinfo {title} {Quantum Computation and Quantum Information: 10th
  Anniversary Edition}}}\ (\bibinfo  {publisher} {Cambridge University Press},\
  \bibinfo {year} {2010})\BibitemShut {NoStop}%
\bibitem [{\citenamefont {Averin}\ \emph {et~al.}(2001)\citenamefont {Averin},
  \citenamefont {Ruggiero},\ and\ \citenamefont
  {Silvestrini}}]{averin2001macroscopic}%
  \BibitemOpen
  \bibfield  {author} {\bibinfo {author} {\bibfnamefont {D.~V.}\ \bibnamefont
  {Averin}}, \bibinfo {author} {\bibfnamefont {B.}~\bibnamefont {Ruggiero}},\
  and\ \bibinfo {author} {\bibfnamefont {P.}~\bibnamefont {Silvestrini}},\
  }\href@noop {} {\emph {\bibinfo {title} {Macroscopic quantum coherence and
  quantum computing}}}\ (\bibinfo  {publisher} {Kluwer Academic/Plenum
  Publishers},\ \bibinfo {address} {New York},\ \bibinfo {year}
  {2001})\BibitemShut {NoStop}%
\bibitem [{\citenamefont {Nitzan}(2006)}]{nitzan06}%
  \BibitemOpen
  \bibfield  {author} {\bibinfo {author} {\bibfnamefont {A.}~\bibnamefont
  {Nitzan}},\ }\href@noop {} {\emph {\bibinfo {title} {Chemical Dynamics in
  Condensed Phases}}}\ (\bibinfo  {publisher} {Oxford University Press},\
  \bibinfo {address} {New York},\ \bibinfo {year} {2006})\BibitemShut {NoStop}%
\bibitem [{\citenamefont {May}\ and\ \citenamefont {K\"{u}hn}(2011)}]{may11}%
  \BibitemOpen
  \bibfield  {author} {\bibinfo {author} {\bibfnamefont {V.}~\bibnamefont
  {May}}\ and\ \bibinfo {author} {\bibfnamefont {O.}~\bibnamefont {K\"{u}hn}},\
  }\href@noop {} {\emph {\bibinfo {title} {Charge and Energy Transfer Dynamics
  in Molecular Systems}}},\ \bibinfo {edition} {3rd}\ ed.\ (\bibinfo
  {publisher} {Wiley-VCH},\ \bibinfo {address} {Weinheim},\ \bibinfo {year}
  {2011})\BibitemShut {NoStop}%
\bibitem [{\citenamefont {Datta}(1995)}]{datta95}%
  \BibitemOpen
  \bibfield  {author} {\bibinfo {author} {\bibfnamefont {S.}~\bibnamefont
  {Datta}},\ }\href@noop {} {\emph {\bibinfo {title} {Electronic Transport in
  Mesoscopic Systems}}}\ (\bibinfo  {publisher} {Cambridge University Press},\
  \bibinfo {address} {New York, NY},\ \bibinfo {year} {1995})\BibitemShut
  {NoStop}%
\bibitem [{\citenamefont {van~der Zant}\ \emph {et~al.}(1991)\citenamefont
  {van~der Zant}, \citenamefont {Geerling}, \citenamefont {Moji},\ and\
  \citenamefont {Kramer}}]{van1991quantum}%
  \BibitemOpen
  \bibfield  {author} {\bibinfo {author} {\bibfnamefont {H.}~\bibnamefont
  {van~der Zant}}, \bibinfo {author} {\bibfnamefont {L.}~\bibnamefont
  {Geerling}}, \bibinfo {author} {\bibfnamefont {J.}~\bibnamefont {Moji}},\
  and\ \bibinfo {author} {\bibfnamefont {B.}~\bibnamefont {Kramer}},\ }\href
  {http://www.springer.com/materials/book/978-0-306-43889-9} {\emph {\bibinfo
  {title} {Quantum Coherence in Mesoscopic Systems ({NATO Science Series B Vol
  254)}}}}\ (\bibinfo  {publisher} {Plenum Press},\ \bibinfo {address} {New
  York},\ \bibinfo {year} {1991})\BibitemShut {NoStop}%
\bibitem [{\citenamefont {Mohseni}\ \emph {et~al.}(2014)\citenamefont
  {Mohseni}, \citenamefont {Omar}, \citenamefont {Engel},\ and\ \citenamefont
  {Plenio}}]{mohseni2014quantum}%
  \BibitemOpen
  \bibfield  {author} {\bibinfo {author} {\bibfnamefont {M.}~\bibnamefont
  {Mohseni}}, \bibinfo {author} {\bibfnamefont {Y.}~\bibnamefont {Omar}},
  \bibinfo {author} {\bibfnamefont {G.~S.}\ \bibnamefont {Engel}},\ and\
  \bibinfo {author} {\bibfnamefont {M.~B.}\ \bibnamefont {Plenio}},\
  }\href@noop {} {\emph {\bibinfo {title} {Quantum effects in biology}}}\
  (\bibinfo  {publisher} {Cambridge University Press},\ \bibinfo {address}
  {Cambridge, England},\ \bibinfo {year} {2014})\BibitemShut {NoStop}%
\bibitem [{\citenamefont {Bulla}\ \emph {et~al.}(2003)\citenamefont {Bulla},
  \citenamefont {Tong},\ and\ \citenamefont {Vojta}}]{bulla2003numerical}%
  \BibitemOpen
  \bibfield  {author} {\bibinfo {author} {\bibfnamefont {R.}~\bibnamefont
  {Bulla}}, \bibinfo {author} {\bibfnamefont {N.-H.}\ \bibnamefont {Tong}},\
  and\ \bibinfo {author} {\bibfnamefont {M.}~\bibnamefont {Vojta}},\ }\bibfield
   {title} {\bibinfo {title} {Numerical renormalization group for bosonic
  systems and application to the sub-ohmic spin-boson model},\ }\href@noop {}
  {\bibfield  {journal} {\bibinfo  {journal} {Phys.~Rev.~Lett.}\ }\textbf
  {\bibinfo {volume} {91}},\ \bibinfo {pages} {170601} (\bibinfo {year}
  {2003})}\BibitemShut {NoStop}%
\bibitem [{\citenamefont {Winter}\ \emph {et~al.}(2009)\citenamefont {Winter},
  \citenamefont {Rieger}, \citenamefont {Vojta},\ and\ \citenamefont
  {Bulla}}]{winter2009quantum}%
  \BibitemOpen
  \bibfield  {author} {\bibinfo {author} {\bibfnamefont {A.}~\bibnamefont
  {Winter}}, \bibinfo {author} {\bibfnamefont {H.}~\bibnamefont {Rieger}},
  \bibinfo {author} {\bibfnamefont {M.}~\bibnamefont {Vojta}},\ and\ \bibinfo
  {author} {\bibfnamefont {R.}~\bibnamefont {Bulla}},\ }\bibfield  {title}
  {\bibinfo {title} {Quantum phase transition in the sub-ohmic spin-boson
  model: {Quantum Monte Carlo} study with a continuous imaginary time cluster
  algorithm},\ }\href@noop {} {\bibfield  {journal} {\bibinfo  {journal}
  {Phys.~Rev.~Lett.}\ }\textbf {\bibinfo {volume} {102}},\ \bibinfo {pages}
  {030601} (\bibinfo {year} {2009})}\BibitemShut {NoStop}%
\bibitem [{\citenamefont {{Spohn}}(1989)}]{spohn89}%
  \BibitemOpen
  \bibfield  {author} {\bibinfo {author} {\bibfnamefont {H.}~\bibnamefont
  {{Spohn}}},\ }\bibfield  {title} {\bibinfo {title} {Ground state(s) of the
  spin-boson hamiltonian},\ }\href@noop {} {\bibfield  {journal} {\bibinfo
  {journal} {Commun. Math. Phys.}\ }\textbf {\bibinfo {volume} {123}},\
  \bibinfo {pages} {277} (\bibinfo {year} {1989})}\BibitemShut {NoStop}%
\bibitem [{\citenamefont {Jang}\ \emph {et~al.}(2008)\citenamefont {Jang},
  \citenamefont {Cheng}, \citenamefont {Reichman},\ and\ \citenamefont
  {Eaves}}]{jang08}%
  \BibitemOpen
  \bibfield  {author} {\bibinfo {author} {\bibfnamefont {S.}~\bibnamefont
  {Jang}}, \bibinfo {author} {\bibfnamefont {Y.-C.}\ \bibnamefont {Cheng}},
  \bibinfo {author} {\bibfnamefont {D.~R.}\ \bibnamefont {Reichman}},\ and\
  \bibinfo {author} {\bibfnamefont {J.~D.}\ \bibnamefont {Eaves}},\ }\bibfield
  {title} {\bibinfo {title} {Theory of coherent resonance energy transfer},\
  }\href@noop {} {\bibfield  {journal} {\bibinfo  {journal} {J.~Chem.~Phys.}\
  }\textbf {\bibinfo {volume} {129}},\ \bibinfo {pages} {101104} (\bibinfo
  {year} {2008})}\BibitemShut {NoStop}%
\bibitem [{\citenamefont {Jang}(2009)}]{jang09}%
  \BibitemOpen
  \bibfield  {author} {\bibinfo {author} {\bibfnamefont {S.}~\bibnamefont
  {Jang}},\ }\bibfield  {title} {\bibinfo {title} {Theory of coherent resonance
  energy transfer for coherent initial condition},\ }\href@noop {} {\bibfield
  {journal} {\bibinfo  {journal} {J.~Chem.~Phys.}\ }\textbf {\bibinfo {volume}
  {131}},\ \bibinfo {pages} {164101} (\bibinfo {year} {2009})}\BibitemShut
  {NoStop}%
\bibitem [{\citenamefont {Palmieri}\ \emph {et~al.}(2009)\citenamefont
  {Palmieri}, \citenamefont {Abramavicius},\ and\ \citenamefont
  {Mukamel}}]{palmieri09}%
  \BibitemOpen
  \bibfield  {author} {\bibinfo {author} {\bibfnamefont {B.}~\bibnamefont
  {Palmieri}}, \bibinfo {author} {\bibfnamefont {D.}~\bibnamefont
  {Abramavicius}},\ and\ \bibinfo {author} {\bibfnamefont {S.}~\bibnamefont
  {Mukamel}},\ }\bibfield  {title} {\bibinfo {title} {Lindblad equations for
  strongly coupled populations and coherences in photosynthetic complexes},\
  }\href@noop {} {\bibfield  {journal} {\bibinfo  {journal} {J.~Chem.~Phys.}\
  }\textbf {\bibinfo {volume} {130}},\ \bibinfo {pages} {204512} (\bibinfo
  {year} {2009})}\BibitemShut {NoStop}%
\bibitem [{\citenamefont {Abramavicius}\ and\ \citenamefont
  {Mukamel}(2010)}]{abramavicius2010quantum}%
  \BibitemOpen
  \bibfield  {author} {\bibinfo {author} {\bibfnamefont {D.}~\bibnamefont
  {Abramavicius}}\ and\ \bibinfo {author} {\bibfnamefont {S.}~\bibnamefont
  {Mukamel}},\ }\bibfield  {title} {\bibinfo {title} {Quantum oscillatory
  exciton migration in photosynthetic reaction centers},\ }\href@noop {}
  {\bibfield  {journal} {\bibinfo  {journal} {J.~Chem.~Phys.}\ }\textbf
  {\bibinfo {volume} {133}},\ \bibinfo {pages} {064510} (\bibinfo {year}
  {2010})}\BibitemShut {NoStop}%
\bibitem [{\citenamefont {Mukamel}(1995)}]{mukamel_book}%
  \BibitemOpen
  \bibfield  {author} {\bibinfo {author} {\bibfnamefont {S.}~\bibnamefont
  {Mukamel}},\ }\href@noop {} {\emph {\bibinfo {title} {Principles of Nonlinear
  Optical Spectroscopy}}}\ (\bibinfo  {publisher} {Oxford University Press},\
  \bibinfo {address} {Oxford},\ \bibinfo {year} {1995})\BibitemShut {NoStop}%
\bibitem [{\citenamefont {M{\"u}hlbacher}\ \emph {et~al.}(2004)\citenamefont
  {M{\"u}hlbacher}, \citenamefont {Ankerhold},\ and\ \citenamefont
  {Escher}}]{muhlbacher2004path}%
  \BibitemOpen
  \bibfield  {author} {\bibinfo {author} {\bibfnamefont {L.}~\bibnamefont
  {M{\"u}hlbacher}}, \bibinfo {author} {\bibfnamefont {J.}~\bibnamefont
  {Ankerhold}},\ and\ \bibinfo {author} {\bibfnamefont {C.}~\bibnamefont
  {Escher}},\ }\bibfield  {title} {\bibinfo {title} {Path-integral {Monte
  Carlo} simulations for electronic dynamics on molecular chains. {I}.
  sequential hopping and super exchange},\ }\href@noop {} {\bibfield  {journal}
  {\bibinfo  {journal} {J.~Chem.~Phys.}\ }\textbf {\bibinfo {volume} {121}},\
  \bibinfo {pages} {12696} (\bibinfo {year} {2004})}\BibitemShut {NoStop}%
\bibitem [{\citenamefont {Thorwart}\ \emph {et~al.}(2009)\citenamefont
  {Thorwart}, \citenamefont {Eckel}, \citenamefont {Reina}, \citenamefont
  {Nalbach},\ and\ \citenamefont {Weiss}}]{thorwart09}%
  \BibitemOpen
  \bibfield  {author} {\bibinfo {author} {\bibfnamefont {M.}~\bibnamefont
  {Thorwart}}, \bibinfo {author} {\bibfnamefont {J.}~\bibnamefont {Eckel}},
  \bibinfo {author} {\bibfnamefont {J.~H.}\ \bibnamefont {Reina}}, \bibinfo
  {author} {\bibfnamefont {P.}~\bibnamefont {Nalbach}},\ and\ \bibinfo {author}
  {\bibfnamefont {S.}~\bibnamefont {Weiss}},\ }\bibfield  {title} {\bibinfo
  {title} {Enhanced quantum entanglement in the {on-Markovian} dynamics of
  biomolecular excitons},\ }\href@noop {} {\bibfield  {journal} {\bibinfo
  {journal} {Chem.~Phys.~Lett.}\ }\textbf {\bibinfo {volume} {478}},\ \bibinfo
  {pages} {234} (\bibinfo {year} {2009})}\BibitemShut {NoStop}%
\bibitem [{\citenamefont {Stockburger}\ and\ \citenamefont
  {Grabert}(2002)}]{stockburger02}%
  \BibitemOpen
  \bibfield  {author} {\bibinfo {author} {\bibfnamefont {J.~T.}\ \bibnamefont
  {Stockburger}}\ and\ \bibinfo {author} {\bibfnamefont {H.}~\bibnamefont
  {Grabert}},\ }\bibfield  {title} {\bibinfo {title} {Exact c-number
  representation of non-markovian quantum dissipation},\ }\href@noop {}
  {\bibfield  {journal} {\bibinfo  {journal} {Phys.~Rev.~Lett.}\ }\textbf
  {\bibinfo {volume} {88}},\ \bibinfo {pages} {170407} (\bibinfo {year}
  {2002})}\BibitemShut {NoStop}%
\bibitem [{\citenamefont {Stockburger}(2004)}]{stockburger04}%
  \BibitemOpen
  \bibfield  {author} {\bibinfo {author} {\bibfnamefont {J.~T.}\ \bibnamefont
  {Stockburger}},\ }\bibfield  {title} {\bibinfo {title} {Simulating spin-boson
  dynamics with stochastic {Liouville-von Neumann} equations},\ }\href@noop {}
  {\bibfield  {journal} {\bibinfo  {journal} {Chem.~Phys.}\ }\textbf {\bibinfo
  {volume} {296}},\ \bibinfo {pages} {159} (\bibinfo {year}
  {2004})}\BibitemShut {NoStop}%
\bibitem [{\citenamefont {Hartmann}\ and\ \citenamefont
  {Strunz}(2017)}]{hartmann2017exact}%
  \BibitemOpen
  \bibfield  {author} {\bibinfo {author} {\bibfnamefont {R.}~\bibnamefont
  {Hartmann}}\ and\ \bibinfo {author} {\bibfnamefont {W.~T.}\ \bibnamefont
  {Strunz}},\ }\bibfield  {title} {\bibinfo {title} {Exact open quantum system
  dynamics using the hierarchy of pure states ({HOPS})},\ }\href@noop {}
  {\bibfield  {journal} {\bibinfo  {journal} {J. Chem. Theory Comput.}\
  }\textbf {\bibinfo {volume} {13}},\ \bibinfo {pages} {5834} (\bibinfo {year}
  {2017})}\BibitemShut {NoStop}%
\bibitem [{\citenamefont {Stockburger}(2016)}]{stock16a}%
  \BibitemOpen
  \bibfield  {author} {\bibinfo {author} {\bibfnamefont {J.~T.}\ \bibnamefont
  {Stockburger}},\ }\bibfield  {title} {\bibinfo {title} {Exact propagation of
  open quantum systems in a system-reservoir context},\ }\href@noop {}
  {\bibfield  {journal} {\bibinfo  {journal} {Europhys. Lett}\ }\textbf
  {\bibinfo {volume} {115}},\ \bibinfo {pages} {40010} (\bibinfo {year}
  {2016})}\BibitemShut {NoStop}%
\bibitem [{\citenamefont {Tanimura}\ and\ \citenamefont
  {Kubo}(1989)}]{tanimura89}%
  \BibitemOpen
  \bibfield  {author} {\bibinfo {author} {\bibfnamefont {Y.}~\bibnamefont
  {Tanimura}}\ and\ \bibinfo {author} {\bibfnamefont {R.}~\bibnamefont
  {Kubo}},\ }\bibfield  {title} {\bibinfo {title} {Time evolution of a quantum
  system in contact with a nearly {Gaussian-Markoffian} noise bath},\
  }\href@noop {} {\bibfield  {journal} {\bibinfo  {journal} {J. Phys. Soc.
  Jpn.}\ }\textbf {\bibinfo {volume} {58}},\ \bibinfo {pages} {101} (\bibinfo
  {year} {1989})}\BibitemShut {NoStop}%
\bibitem [{\citenamefont {Tanimura}\ and\ \citenamefont
  {Mukamel}(1993)}]{tanimura1993two}%
  \BibitemOpen
  \bibfield  {author} {\bibinfo {author} {\bibfnamefont {Y.}~\bibnamefont
  {Tanimura}}\ and\ \bibinfo {author} {\bibfnamefont {S.}~\bibnamefont
  {Mukamel}},\ }\bibfield  {title} {\bibinfo {title} {Two-dimensional
  femtosecond vibrational spectroscopy of liquids},\ }\href@noop {} {\bibfield
  {journal} {\bibinfo  {journal} {J.~Chem.~Phys.}\ }\textbf {\bibinfo {volume}
  {99}},\ \bibinfo {pages} {9496} (\bibinfo {year} {1993})}\BibitemShut
  {NoStop}%
\bibitem [{\citenamefont {Tanimura}(2006)}]{tanimura06}%
  \BibitemOpen
  \bibfield  {author} {\bibinfo {author} {\bibfnamefont {Y.}~\bibnamefont
  {Tanimura}},\ }\bibfield  {title} {\bibinfo {title} {{Stochastic Liouville,
  Langevin, Fokker-Planck}, and master equation approaches to quantum
  dissipative systems},\ }\href@noop {} {\bibfield  {journal} {\bibinfo
  {journal} {J. Phys. Soc. Jpn.}\ }\textbf {\bibinfo {volume} {75}},\ \bibinfo
  {pages} {082001} (\bibinfo {year} {2006})}\BibitemShut {NoStop}%
\bibitem [{\citenamefont {Abramavicius}\ \emph {et~al.}(2009)\citenamefont
  {Abramavicius}, \citenamefont {Palmieri}, \citenamefont {Voronine},
  \citenamefont {Sanda},\ and\ \citenamefont
  {Mukamel}}]{abramavicius2009coherent}%
  \BibitemOpen
  \bibfield  {author} {\bibinfo {author} {\bibfnamefont {D.}~\bibnamefont
  {Abramavicius}}, \bibinfo {author} {\bibfnamefont {B.}~\bibnamefont
  {Palmieri}}, \bibinfo {author} {\bibfnamefont {D.~V.}\ \bibnamefont
  {Voronine}}, \bibinfo {author} {\bibfnamefont {F.}~\bibnamefont {Sanda}},\
  and\ \bibinfo {author} {\bibfnamefont {S.}~\bibnamefont {Mukamel}},\
  }\bibfield  {title} {\bibinfo {title} {Coherent multidimensional optical
  spectroscopy of excitons in molecular aggregates; quasiparticle versus
  supermolecule perspectives},\ }\href@noop {} {\bibfield  {journal} {\bibinfo
  {journal} {Chem. Rev.}\ }\textbf {\bibinfo {volume} {109}},\ \bibinfo {pages}
  {2350} (\bibinfo {year} {2009})}\BibitemShut {NoStop}%
\bibitem [{\citenamefont {Ishizaki}\ and\ \citenamefont
  {Fleming}(2009)}]{ishizaki09b}%
  \BibitemOpen
  \bibfield  {author} {\bibinfo {author} {\bibfnamefont {A.}~\bibnamefont
  {Ishizaki}}\ and\ \bibinfo {author} {\bibfnamefont {G.~R.}\ \bibnamefont
  {Fleming}},\ }\bibfield  {title} {\bibinfo {title} {Theoretical examination
  of quantum coherence in a photosynthetic system at physiological
  temperature},\ }\href@noop {} {\bibfield  {journal} {\bibinfo  {journal}
  {Proc. Natl. Acad. Sci. U.S.A.}\ }\textbf {\bibinfo {volume} {106}},\
  \bibinfo {pages} {17255} (\bibinfo {year} {2009})}\BibitemShut {NoStop}%
\bibitem [{\citenamefont {Sarovar}\ and\ \citenamefont
  {Grace}(2012)}]{sarovar2012reduced}%
  \BibitemOpen
  \bibfield  {author} {\bibinfo {author} {\bibfnamefont {M.}~\bibnamefont
  {Sarovar}}\ and\ \bibinfo {author} {\bibfnamefont {M.~D.}\ \bibnamefont
  {Grace}},\ }\bibfield  {title} {\bibinfo {title} {Reduced equations of motion
  for quantum systems driven by diffusive markov processes},\ }\href@noop {}
  {\bibfield  {journal} {\bibinfo  {journal} {Phys.~Rev.~Lett.}\ }\textbf
  {\bibinfo {volume} {109}},\ \bibinfo {pages} {130401} (\bibinfo {year}
  {2012})}\BibitemShut {NoStop}%
\bibitem [{\citenamefont {Yan}(2014)}]{yan2014theory}%
  \BibitemOpen
  \bibfield  {author} {\bibinfo {author} {\bibfnamefont {Y.}~\bibnamefont
  {Yan}},\ }\bibfield  {title} {\bibinfo {title} {{Theory of open quantum
  systems with bath of electrons and phonons and spins: Many-dissipaton density
  matrixes approach}},\ }\href@noop {} {\bibfield  {journal} {\bibinfo
  {journal} {J.~Chem.~Phys.}\ }\textbf {\bibinfo {volume} {140}},\ \bibinfo
  {pages} {054105} (\bibinfo {year} {2014})}\BibitemShut {NoStop}%
\bibitem [{\citenamefont {Tanimura}(2020)}]{tanimura2020numerically}%
  \BibitemOpen
  \bibfield  {author} {\bibinfo {author} {\bibfnamefont {Y.}~\bibnamefont
  {Tanimura}},\ }\bibfield  {title} {\bibinfo {title} {Numerically “exact”
  approach to open quantum dynamics: The hierarchical equations of motion
  ({HEOM})},\ }\href@noop {} {\bibfield  {journal} {\bibinfo  {journal}
  {J.~Chem.~Phys.}\ }\textbf {\bibinfo {volume} {153}},\ \bibinfo {pages}
  {020901} (\bibinfo {year} {2020})}\BibitemShut {NoStop}%
\bibitem [{rem()}]{remark-struc}%
  \BibitemOpen
  \href@noop {} {\bibinfo {title} {When speaking of strucured reservoirs, we
  consider the case of continuos spectra with pronounced, but finite-width
  peaks. {More} suitable methods exist for truly discrete
  reservoirs.}}\BibitemShut {Stop}%
\bibitem [{\citenamefont {Makri}\ and\ \citenamefont
  {Makarov}(1995)}]{makri1995tensor}%
  \BibitemOpen
  \bibfield  {author} {\bibinfo {author} {\bibfnamefont {N.}~\bibnamefont
  {Makri}}\ and\ \bibinfo {author} {\bibfnamefont {D.~E.}\ \bibnamefont
  {Makarov}},\ }\bibfield  {title} {\bibinfo {title} {Tensor propagator for
  iterative quantum time evolution of reduced density matrices. i. theory},\
  }\href@noop {} {\bibfield  {journal} {\bibinfo  {journal} {J.~Chem.~Phys.}\
  }\textbf {\bibinfo {volume} {102}},\ \bibinfo {pages} {4600} (\bibinfo {year}
  {1995})}\BibitemShut {NoStop}%
\bibitem [{\citenamefont {Shao}(2004)}]{shao2004decoupling}%
  \BibitemOpen
  \bibfield  {author} {\bibinfo {author} {\bibfnamefont {J.}~\bibnamefont
  {Shao}},\ }\bibfield  {title} {\bibinfo {title} {Decoupling quantum
  dissipation interaction via stochastic fields},\ }\href@noop {} {\bibfield
  {journal} {\bibinfo  {journal} {J.~Chem.~Phys.}\ }\textbf {\bibinfo {volume}
  {120}},\ \bibinfo {pages} {5053} (\bibinfo {year} {2004})}\BibitemShut
  {NoStop}%
\bibitem [{\citenamefont {Cohen}\ \emph {et~al.}(2015)\citenamefont {Cohen},
  \citenamefont {Gull}, \citenamefont {Reichman},\ and\ \citenamefont
  {Millis}}]{cohen2005taming}%
  \BibitemOpen
  \bibfield  {author} {\bibinfo {author} {\bibfnamefont {G.}~\bibnamefont
  {Cohen}}, \bibinfo {author} {\bibfnamefont {E.}~\bibnamefont {Gull}},
  \bibinfo {author} {\bibfnamefont {D.~R.}\ \bibnamefont {Reichman}},\ and\
  \bibinfo {author} {\bibfnamefont {A.~J.}\ \bibnamefont {Millis}},\ }\bibfield
   {title} {\bibinfo {title} {Taming the dynamical sign problem in real-time
  evolution of quantum many-body problems},\ }\href
  {https://doi.org/10.1103/PhysRevLett.115.266802} {\bibfield  {journal}
  {\bibinfo  {journal} {Phys. Rev. Lett.}\ }\textbf {\bibinfo {volume} {115}},\
  \bibinfo {pages} {266802} (\bibinfo {year} {2015})}\BibitemShut {NoStop}%
\bibitem [{\citenamefont {Kast}\ and\ \citenamefont
  {Ankerhold}(2013)}]{kast2013persistence}%
  \BibitemOpen
  \bibfield  {author} {\bibinfo {author} {\bibfnamefont {D.}~\bibnamefont
  {Kast}}\ and\ \bibinfo {author} {\bibfnamefont {J.}~\bibnamefont
  {Ankerhold}},\ }\bibfield  {title} {\bibinfo {title} {Persistence of coherent
  quantum dynamics at strong dissipation},\ }\href
  {https://doi.org/10.1103/PhysRevLett.110.010402} {\bibfield  {journal}
  {\bibinfo  {journal} {Phys. Rev. Lett.}\ }\textbf {\bibinfo {volume} {110}},\
  \bibinfo {pages} {010402} (\bibinfo {year} {2013})}\BibitemShut {NoStop}%
\bibitem [{\citenamefont {Suess}\ \emph {et~al.}(2014)\citenamefont {Suess},
  \citenamefont {Eisfeld},\ and\ \citenamefont {Strunz}}]{suess2014hierarchy}%
  \BibitemOpen
  \bibfield  {author} {\bibinfo {author} {\bibfnamefont {D.}~\bibnamefont
  {Suess}}, \bibinfo {author} {\bibfnamefont {A.}~\bibnamefont {Eisfeld}},\
  and\ \bibinfo {author} {\bibfnamefont {W.~T.}\ \bibnamefont {Strunz}},\
  }\bibfield  {title} {\bibinfo {title} {Hierarchy of stochastic pure states
  for open quantum system dynamics},\ }\href@noop {} {\bibfield  {journal}
  {\bibinfo  {journal} {Phys.~Rev.~Lett.}\ }\textbf {\bibinfo {volume} {113}},\
  \bibinfo {pages} {150403} (\bibinfo {year} {2014})}\BibitemShut {NoStop}%
\bibitem [{\citenamefont {Tamascelli}\ \emph {et~al.}(2018)\citenamefont
  {Tamascelli}, \citenamefont {Smirne}, \citenamefont {Huelga},\ and\
  \citenamefont {Plenio}}]{tamascelli2018nonperturbative}%
  \BibitemOpen
  \bibfield  {author} {\bibinfo {author} {\bibfnamefont {D.}~\bibnamefont
  {Tamascelli}}, \bibinfo {author} {\bibfnamefont {A.}~\bibnamefont {Smirne}},
  \bibinfo {author} {\bibfnamefont {S.~F.}\ \bibnamefont {Huelga}},\ and\
  \bibinfo {author} {\bibfnamefont {M.~B.}\ \bibnamefont {Plenio}},\ }\bibfield
   {title} {\bibinfo {title} {Nonperturbative treatment of non-markovian
  dynamics of open quantum systems},\ }\href@noop {} {\bibfield  {journal}
  {\bibinfo  {journal} {Phys.~Rev.~Lett.}\ }\textbf {\bibinfo {volume} {120}},\
  \bibinfo {pages} {030402} (\bibinfo {year} {2018})}\BibitemShut {NoStop}%
\bibitem [{\citenamefont {Cygorek}\ \emph {et~al.}(2017)\citenamefont
  {Cygorek}, \citenamefont {Barth}, \citenamefont {Ungar}, \citenamefont
  {Vagov},\ and\ \citenamefont {Axt}}]{cygorek2017nonlinear}%
  \BibitemOpen
  \bibfield  {author} {\bibinfo {author} {\bibfnamefont {M.}~\bibnamefont
  {Cygorek}}, \bibinfo {author} {\bibfnamefont {A.~M.}\ \bibnamefont {Barth}},
  \bibinfo {author} {\bibfnamefont {F.}~\bibnamefont {Ungar}}, \bibinfo
  {author} {\bibfnamefont {A.}~\bibnamefont {Vagov}},\ and\ \bibinfo {author}
  {\bibfnamefont {V.~M.}\ \bibnamefont {Axt}},\ }\bibfield  {title} {\bibinfo
  {title} {Nonlinear cavity feeding and unconventional photon statistics in
  solid-state cavity qed revealed by many-level real-time path-integral
  calculations},\ }\href {https://doi.org/10.1103/PhysRevB.96.201201}
  {\bibfield  {journal} {\bibinfo  {journal} {Phys. Rev. B}\ }\textbf {\bibinfo
  {volume} {96}},\ \bibinfo {pages} {201201(R)} (\bibinfo {year}
  {2017})}\BibitemShut {NoStop}%
\bibitem [{\citenamefont {Strathearn}\ \emph {et~al.}(2018)\citenamefont
  {Strathearn}, \citenamefont {Kirton}, \citenamefont {Kilda}, \citenamefont
  {Keeling},\ and\ \citenamefont {Lovett}}]{strathearn2018efficient}%
  \BibitemOpen
  \bibfield  {author} {\bibinfo {author} {\bibfnamefont {A.}~\bibnamefont
  {Strathearn}}, \bibinfo {author} {\bibfnamefont {P.}~\bibnamefont {Kirton}},
  \bibinfo {author} {\bibfnamefont {D.}~\bibnamefont {Kilda}}, \bibinfo
  {author} {\bibfnamefont {J.}~\bibnamefont {Keeling}},\ and\ \bibinfo {author}
  {\bibfnamefont {B.~W.}\ \bibnamefont {Lovett}},\ }\bibfield  {title}
  {\bibinfo {title} {Efficient non-markovian quantum dynamics using
  time-evolving matrix product operators},\ }\href@noop {} {\bibfield
  {journal} {\bibinfo  {journal} {Nat. Commun.}\ }\textbf {\bibinfo {volume}
  {9}},\ \bibinfo {pages} {3322} (\bibinfo {year} {2018})}\BibitemShut
  {NoStop}%
\bibitem [{\citenamefont {Pollock}\ \emph {et~al.}(2018)\citenamefont
  {Pollock}, \citenamefont {Rodr{\'\i}guez-Rosario}, \citenamefont
  {Frauenheim}, \citenamefont {Paternostro},\ and\ \citenamefont
  {Modi}}]{pollock2018non}%
  \BibitemOpen
  \bibfield  {author} {\bibinfo {author} {\bibfnamefont {F.~A.}\ \bibnamefont
  {Pollock}}, \bibinfo {author} {\bibfnamefont {C.}~\bibnamefont
  {Rodr{\'\i}guez-Rosario}}, \bibinfo {author} {\bibfnamefont {T.}~\bibnamefont
  {Frauenheim}}, \bibinfo {author} {\bibfnamefont {M.}~\bibnamefont
  {Paternostro}},\ and\ \bibinfo {author} {\bibfnamefont {K.}~\bibnamefont
  {Modi}},\ }\bibfield  {title} {\bibinfo {title} {Non-markovian quantum
  processes: Complete framework and efficient characterization},\ }\href@noop
  {} {\bibfield  {journal} {\bibinfo  {journal} {Phys.~Rev.~A}\ }\textbf
  {\bibinfo {volume} {97}},\ \bibinfo {pages} {012127} (\bibinfo {year}
  {2018})}\BibitemShut {NoStop}%
\bibitem [{\citenamefont {Bose}\ and\ \citenamefont
  {Walters}(2022)}]{bose2022multisite}%
  \BibitemOpen
  \bibfield  {author} {\bibinfo {author} {\bibfnamefont {A.}~\bibnamefont
  {Bose}}\ and\ \bibinfo {author} {\bibfnamefont {P.~L.}\ \bibnamefont
  {Walters}},\ }\bibfield  {title} {\bibinfo {title} {A multisite decomposition
  of the tensor network path integrals},\ }\href@noop {} {\bibfield  {journal}
  {\bibinfo  {journal} {J.~Chem.~Phys.}\ }\textbf {\bibinfo {volume} {156}},\
  \bibinfo {pages} {024101} (\bibinfo {year} {2022})}\BibitemShut {NoStop}%
\bibitem [{\citenamefont {Cygorek}\ \emph {et~al.}(2022)\citenamefont
  {Cygorek}, \citenamefont {Cosacchi}, \citenamefont {Vagov}, \citenamefont
  {Axt}, \citenamefont {Lovett}, \citenamefont {Keeling},\ and\ \citenamefont
  {Gauger}}]{cygorek2022simulation}%
  \BibitemOpen
  \bibfield  {author} {\bibinfo {author} {\bibfnamefont {M.}~\bibnamefont
  {Cygorek}}, \bibinfo {author} {\bibfnamefont {M.}~\bibnamefont {Cosacchi}},
  \bibinfo {author} {\bibfnamefont {A.}~\bibnamefont {Vagov}}, \bibinfo
  {author} {\bibfnamefont {V.~M.}\ \bibnamefont {Axt}}, \bibinfo {author}
  {\bibfnamefont {B.~W.}\ \bibnamefont {Lovett}}, \bibinfo {author}
  {\bibfnamefont {J.}~\bibnamefont {Keeling}},\ and\ \bibinfo {author}
  {\bibfnamefont {E.~M.}\ \bibnamefont {Gauger}},\ }\bibfield  {title}
  {\bibinfo {title} {Simulation of open quantum systems by automated
  compression of arbitrary environments},\ }\href@noop {} {\bibfield  {journal}
  {\bibinfo  {journal} {Nat. Phys.}\ }\textbf {\bibinfo {volume} {18}},\
  \bibinfo {pages} {662} (\bibinfo {year} {2022})}\BibitemShut {NoStop}%
\bibitem [{\citenamefont {Xu}\ \emph {et~al.}(2005)\citenamefont {Xu},
  \citenamefont {Cui}, \citenamefont {Li}, \citenamefont {Mo},\ and\
  \citenamefont {Yan}}]{xu05}%
  \BibitemOpen
  \bibfield  {author} {\bibinfo {author} {\bibfnamefont {R.-X.}\ \bibnamefont
  {Xu}}, \bibinfo {author} {\bibfnamefont {P.}~\bibnamefont {Cui}}, \bibinfo
  {author} {\bibfnamefont {X.-Q.}\ \bibnamefont {Li}}, \bibinfo {author}
  {\bibfnamefont {Y.}~\bibnamefont {Mo}},\ and\ \bibinfo {author}
  {\bibfnamefont {Y.-J.}\ \bibnamefont {Yan}},\ }\bibfield  {title} {\bibinfo
  {title} {Exact quantum master equation via the calculus on path integrals},\
  }\href@noop {} {\bibfield  {journal} {\bibinfo  {journal} {J.~Chem.~Phys.}\
  }\textbf {\bibinfo {volume} {122}},\ \bibinfo {pages} {041103} (\bibinfo
  {year} {2005})}\BibitemShut {NoStop}%
\bibitem [{\citenamefont {Duan}\ \emph {et~al.}(2017)\citenamefont {Duan},
  \citenamefont {Tang}, \citenamefont {Cao},\ and\ \citenamefont
  {Wu}}]{duan17}%
  \BibitemOpen
  \bibfield  {author} {\bibinfo {author} {\bibfnamefont {C.}~\bibnamefont
  {Duan}}, \bibinfo {author} {\bibfnamefont {Z.}~\bibnamefont {Tang}}, \bibinfo
  {author} {\bibfnamefont {J.}~\bibnamefont {Cao}},\ and\ \bibinfo {author}
  {\bibfnamefont {J.}~\bibnamefont {Wu}},\ }\bibfield  {title} {\bibinfo
  {title} {Zero-temperature localization in a sub-ohmic spin-boson model
  investigated by an extended hierarchy equation of motion},\ }\href@noop {}
  {\bibfield  {journal} {\bibinfo  {journal} {Phys.~Rev.~B}\ }\textbf {\bibinfo
  {volume} {95}},\ \bibinfo {pages} {214308} (\bibinfo {year}
  {2017})}\BibitemShut {NoStop}%
\bibitem [{\citenamefont {Rahman}\ and\ \citenamefont
  {Kleinekath{\"o}fer}(2019)}]{rahman2019chebyshev}%
  \BibitemOpen
  \bibfield  {author} {\bibinfo {author} {\bibfnamefont {H.}~\bibnamefont
  {Rahman}}\ and\ \bibinfo {author} {\bibfnamefont {U.}~\bibnamefont
  {Kleinekath{\"o}fer}},\ }\bibfield  {title} {\bibinfo {title} {Chebyshev
  hierarchical equations of motion for systems with arbitrary spectral
  densities and temperatures},\ }\href@noop {} {\bibfield  {journal} {\bibinfo
  {journal} {J.~Chem.~Phys.}\ }\textbf {\bibinfo {volume} {150}},\ \bibinfo
  {pages} {244104} (\bibinfo {year} {2019})}\BibitemShut {NoStop}%
\bibitem [{\citenamefont {Cui}\ \emph {et~al.}(2019)\citenamefont {Cui},
  \citenamefont {Zhang}, \citenamefont {Zheng}, \citenamefont {Xu},\ and\
  \citenamefont {Yan}}]{cui2019highly}%
  \BibitemOpen
  \bibfield  {author} {\bibinfo {author} {\bibfnamefont {L.}~\bibnamefont
  {Cui}}, \bibinfo {author} {\bibfnamefont {H.-D.}\ \bibnamefont {Zhang}},
  \bibinfo {author} {\bibfnamefont {X.}~\bibnamefont {Zheng}}, \bibinfo
  {author} {\bibfnamefont {R.-X.}\ \bibnamefont {Xu}},\ and\ \bibinfo {author}
  {\bibfnamefont {Y.}~\bibnamefont {Yan}},\ }\bibfield  {title} {\bibinfo
  {title} {Highly efficient and accurate sum-over-poles expansion of fermi and
  bose functions at near zero temperatures: Fano spectrum decomposition
  scheme},\ }\href@noop {} {\bibfield  {journal} {\bibinfo  {journal}
  {J.~Chem.~Phys.}\ }\textbf {\bibinfo {volume} {151}},\ \bibinfo {pages}
  {024110} (\bibinfo {year} {2019})}\BibitemShut {NoStop}%
\bibitem [{\citenamefont {Ikeda}\ and\ \citenamefont
  {Scholes}(2020)}]{ikeda2020generalization}%
  \BibitemOpen
  \bibfield  {author} {\bibinfo {author} {\bibfnamefont {T.}~\bibnamefont
  {Ikeda}}\ and\ \bibinfo {author} {\bibfnamefont {G.~D.}\ \bibnamefont
  {Scholes}},\ }\bibfield  {title} {\bibinfo {title} {Generalization of the
  hierarchical equations of motion theory for efficient calculations with
  arbitrary correlation functions},\ }\href@noop {} {\bibfield  {journal}
  {\bibinfo  {journal} {J.~Chem.~Phys.}\ }\textbf {\bibinfo {volume} {152}},\
  \bibinfo {pages} {204101} (\bibinfo {year} {2020})}\BibitemShut {NoStop}%
\bibitem [{\citenamefont {Chen}\ \emph {et~al.}(2022)\citenamefont {Chen},
  \citenamefont {Wang}, \citenamefont {Zheng}, \citenamefont {Xu},\ and\
  \citenamefont {Yan}}]{chen2022universal}%
  \BibitemOpen
  \bibfield  {author} {\bibinfo {author} {\bibfnamefont {Z.-H.}\ \bibnamefont
  {Chen}}, \bibinfo {author} {\bibfnamefont {Y.}~\bibnamefont {Wang}}, \bibinfo
  {author} {\bibfnamefont {X.}~\bibnamefont {Zheng}}, \bibinfo {author}
  {\bibfnamefont {R.-X.}\ \bibnamefont {Xu}},\ and\ \bibinfo {author}
  {\bibfnamefont {Y.}~\bibnamefont {Yan}},\ }\bibfield  {title} {\bibinfo
  {title} {{Universal time-domain Prony fitting decomposition for optimized
  hierarchical quantum master equations}},\ }\href@noop {} {\bibfield
  {journal} {\bibinfo  {journal} {J.~Chem.~Phys.}\ }\textbf {\bibinfo {volume}
  {156}},\ \bibinfo {pages} {221102} (\bibinfo {year} {2022})}\BibitemShut
  {NoStop}%
\bibitem [{\citenamefont {Shi}\ \emph {et~al.}(2018)\citenamefont {Shi},
  \citenamefont {Xu}, \citenamefont {Yan},\ and\ \citenamefont
  {Xu}}]{shi2018efficient}%
  \BibitemOpen
  \bibfield  {author} {\bibinfo {author} {\bibfnamefont {Q.}~\bibnamefont
  {Shi}}, \bibinfo {author} {\bibfnamefont {Y.}~\bibnamefont {Xu}}, \bibinfo
  {author} {\bibfnamefont {Y.}~\bibnamefont {Yan}},\ and\ \bibinfo {author}
  {\bibfnamefont {M.}~\bibnamefont {Xu}},\ }\bibfield  {title} {\bibinfo
  {title} {Efficient propagation of the hierarchical equations of motion using
  the matrix product state method},\ }\href@noop {} {\bibfield  {journal}
  {\bibinfo  {journal} {J.~Chem.~Phys.}\ }\textbf {\bibinfo {volume} {148}},\
  \bibinfo {pages} {174102} (\bibinfo {year} {2018})}\BibitemShut {NoStop}%
\bibitem [{\citenamefont {Borrelli}(2019)}]{borrelli2019density}%
  \BibitemOpen
  \bibfield  {author} {\bibinfo {author} {\bibfnamefont {R.}~\bibnamefont
  {Borrelli}},\ }\bibfield  {title} {\bibinfo {title} {Density matrix dynamics
  in twin-formulation: An efficient methodology based on tensor-train
  representation of reduced equations of motion},\ }\href@noop {} {\bibfield
  {journal} {\bibinfo  {journal} {J.~Chem.~Phys.}\ }\textbf {\bibinfo {volume}
  {150}},\ \bibinfo {pages} {234102} (\bibinfo {year} {2019})}\BibitemShut
  {NoStop}%
\bibitem [{\citenamefont {Yan}\ \emph {et~al.}(2021)\citenamefont {Yan},
  \citenamefont {Xu}, \citenamefont {Li},\ and\ \citenamefont
  {Shi}}]{yan2021efficient}%
  \BibitemOpen
  \bibfield  {author} {\bibinfo {author} {\bibfnamefont {Y.}~\bibnamefont
  {Yan}}, \bibinfo {author} {\bibfnamefont {M.}~\bibnamefont {Xu}}, \bibinfo
  {author} {\bibfnamefont {T.}~\bibnamefont {Li}},\ and\ \bibinfo {author}
  {\bibfnamefont {Q.}~\bibnamefont {Shi}},\ }\bibfield  {title} {\bibinfo
  {title} {Efficient propagation of the hierarchical equations of motion using
  the tucker and hierarchical tucker tensors},\ }\href@noop {} {\bibfield
  {journal} {\bibinfo  {journal} {J.~Chem.~Phys.}\ }\textbf {\bibinfo {volume}
  {154}},\ \bibinfo {pages} {194104} (\bibinfo {year} {2021})}\BibitemShut
  {NoStop}%
\bibitem [{\citenamefont {Croy}\ and\ \citenamefont {Saalmann}(2009)}]{croy09}%
  \BibitemOpen
  \bibfield  {author} {\bibinfo {author} {\bibfnamefont {A.}~\bibnamefont
  {Croy}}\ and\ \bibinfo {author} {\bibfnamefont {U.}~\bibnamefont
  {Saalmann}},\ }\bibfield  {title} {\bibinfo {title} {Propagation scheme for
  nonequilibrium dynamics of electron transport in nanoscale devices},\
  }\href@noop {} {\bibfield  {journal} {\bibinfo  {journal} {Phys.~Rev.~B}\
  }\textbf {\bibinfo {volume} {80}},\ \bibinfo {pages} {245311} (\bibinfo
  {year} {2009})}\BibitemShut {NoStop}%
\bibitem [{\citenamefont {Beach}\ \emph {et~al.}(2000)\citenamefont {Beach},
  \citenamefont {Gooding},\ and\ \citenamefont
  {Marsiglio}}]{beach2000reliable}%
  \BibitemOpen
  \bibfield  {author} {\bibinfo {author} {\bibfnamefont {K.~S.~D.}\
  \bibnamefont {Beach}}, \bibinfo {author} {\bibfnamefont {R.~J.}\ \bibnamefont
  {Gooding}},\ and\ \bibinfo {author} {\bibfnamefont {F.}~\bibnamefont
  {Marsiglio}},\ }\bibfield  {title} {\bibinfo {title} {Reliable pad\'e
  analytical continuation method based on a high-accuracy symbolic computation
  algorithm},\ }\href {https://doi.org/10.1103/PhysRevB.61.5147} {\bibfield
  {journal} {\bibinfo  {journal} {Phys. Rev. B}\ }\textbf {\bibinfo {volume}
  {61}},\ \bibinfo {pages} {5147} (\bibinfo {year} {2000})}\BibitemShut
  {NoStop}%
\bibitem [{\citenamefont {Gu}\ \emph {et~al.}(2020)\citenamefont {Gu},
  \citenamefont {Chen}, \citenamefont {Wang},\ and\ \citenamefont
  {Zhang}}]{gu2020generalized}%
  \BibitemOpen
  \bibfield  {author} {\bibinfo {author} {\bibfnamefont {J.}~\bibnamefont
  {Gu}}, \bibinfo {author} {\bibfnamefont {J.}~\bibnamefont {Chen}}, \bibinfo
  {author} {\bibfnamefont {Y.}~\bibnamefont {Wang}},\ and\ \bibinfo {author}
  {\bibfnamefont {X.-G.}\ \bibnamefont {Zhang}},\ }\bibfield  {title} {\bibinfo
  {title} {Generalized quadrature for finite temperature green’s function
  methods},\ }\href@noop {} {\bibfield  {journal} {\bibinfo  {journal} {Comput.
  Phys. Commun.}\ }\textbf {\bibinfo {volume} {253}},\ \bibinfo {pages}
  {107178} (\bibinfo {year} {2020})}\BibitemShut {NoStop}%
\bibitem [{\citenamefont {Pleasance}\ \emph {et~al.}(2020)\citenamefont
  {Pleasance}, \citenamefont {Garraway},\ and\ \citenamefont
  {Petruccione}}]{pleasance2020generalized}%
  \BibitemOpen
  \bibfield  {author} {\bibinfo {author} {\bibfnamefont {G.}~\bibnamefont
  {Pleasance}}, \bibinfo {author} {\bibfnamefont {B.~M.}\ \bibnamefont
  {Garraway}},\ and\ \bibinfo {author} {\bibfnamefont {F.}~\bibnamefont
  {Petruccione}},\ }\bibfield  {title} {\bibinfo {title} {Generalized theory of
  pseudomodes for exact descriptions of {non-Markovian} quantum processes},\
  }\href {https://doi.org/10.1103/PhysRevResearch.2.043058} {\bibfield
  {journal} {\bibinfo  {journal} {Phys. Rev. Res.}\ }\textbf {\bibinfo {volume}
  {2}},\ \bibinfo {pages} {043058} (\bibinfo {year} {2020})}\BibitemShut
  {NoStop}%
\bibitem [{\citenamefont {Trivedi}\ \emph {et~al.}(2021)\citenamefont
  {Trivedi}, \citenamefont {Malz},\ and\ \citenamefont
  {Cirac}}]{trivedi2021convergence}%
  \BibitemOpen
  \bibfield  {author} {\bibinfo {author} {\bibfnamefont {R.}~\bibnamefont
  {Trivedi}}, \bibinfo {author} {\bibfnamefont {D.}~\bibnamefont {Malz}},\ and\
  \bibinfo {author} {\bibfnamefont {J.~I.}\ \bibnamefont {Cirac}},\ }\bibfield
  {title} {\bibinfo {title} {Convergence guarantees for discrete mode
  approximations to non-markovian quantum baths},\ }\href@noop {} {\bibfield
  {journal} {\bibinfo  {journal} {Phys.~Rev.~Lett.}\ }\textbf {\bibinfo
  {volume} {127}},\ \bibinfo {pages} {250404} (\bibinfo {year}
  {2021})}\BibitemShut {NoStop}%
\bibitem [{\citenamefont {Awad}\ and\ \citenamefont {Khanna}(2015)}]{Awad2015}%
  \BibitemOpen
  \bibfield  {author} {\bibinfo {author} {\bibfnamefont {M.}~\bibnamefont
  {Awad}}\ and\ \bibinfo {author} {\bibfnamefont {R.}~\bibnamefont {Khanna}},\
  }\bibinfo {title} {Hidden markov model},\ in\ \href
  {https://doi.org/10.1007/978-1-4302-5990-9_5} {\emph {\bibinfo {booktitle}
  {Efficient Learning Machines: Theories, Concepts, and Applications for
  Engineers and System Designers}}}\ (\bibinfo  {publisher} {Apress},\ \bibinfo
  {address} {Berkeley, CA},\ \bibinfo {year} {2015})\ pp.\ \bibinfo {pages}
  {81--104}\BibitemShut {NoStop}%
\bibitem [{\citenamefont {Shiba}(1975)}]{shiba75a}%
  \BibitemOpen
  \bibfield  {author} {\bibinfo {author} {\bibfnamefont {H.}~\bibnamefont
  {Shiba}},\ }\bibfield  {title} {\bibinfo {title} {The {Korringa} relation for
  the impurity nuclear spin-lattice relaxation in dilute kondo alloys},\
  }\href@noop {} {\bibfield  {journal} {\bibinfo  {journal} {Prog. Theor.
  Phys.}\ }\textbf {\bibinfo {volume} {54}},\ \bibinfo {pages} {967} (\bibinfo
  {year} {1975})}\BibitemShut {NoStop}%
\bibitem [{\citenamefont {Sassetti}\ and\ \citenamefont
  {Weiss}(1990)}]{sasse90b}%
  \BibitemOpen
  \bibfield  {author} {\bibinfo {author} {\bibfnamefont {M.}~\bibnamefont
  {Sassetti}}\ and\ \bibinfo {author} {\bibfnamefont {U.}~\bibnamefont
  {Weiss}},\ }\bibfield  {title} {\bibinfo {title} {Universality in the
  dissipative two-state system},\ }\href
  {https://doi.org/http://dx.doi.org/10.1103/PhysRevLett.65.2262} {\bibfield
  {journal} {\bibinfo  {journal} {Phys. Rev. Lett.}\ }\textbf {\bibinfo
  {volume} {65}},\ \bibinfo {pages} {2262} (\bibinfo {year}
  {1990})}\BibitemShut {NoStop}%
\bibitem [{Note1()}]{Note1}%
  \BibitemOpen
  \bibinfo {note} {The canonical structure has no pole at $\omega
  =0$.}\BibitemShut {Stop}%
\bibitem [{SI()}]{SI}%
  \BibitemOpen
  \href@noop {} {\bibinfo {title} {See supplemental material at [url] for
  further information, which includes {Refs}.
  \cite{feynman63,echave92,lubich15,haegeman2016unifying}}}\BibitemShut
  {NoStop}%
\bibitem [{\citenamefont {Shi}\ \emph {et~al.}(2009)\citenamefont {Shi},
  \citenamefont {Chen}, \citenamefont {Nan}, \citenamefont {Xu},\ and\
  \citenamefont {Yan}}]{shi2009efficient}%
  \BibitemOpen
  \bibfield  {author} {\bibinfo {author} {\bibfnamefont {Q.}~\bibnamefont
  {Shi}}, \bibinfo {author} {\bibfnamefont {L.}~\bibnamefont {Chen}}, \bibinfo
  {author} {\bibfnamefont {G.}~\bibnamefont {Nan}}, \bibinfo {author}
  {\bibfnamefont {R.-X.}\ \bibnamefont {Xu}},\ and\ \bibinfo {author}
  {\bibfnamefont {Y.}~\bibnamefont {Yan}},\ }\bibfield  {title} {\bibinfo
  {title} {Efficient hierarchical {Liouville} space propagator to quantum
  dissipative dynamics},\ }\href@noop {} {\bibfield  {journal} {\bibinfo
  {journal} {J.~Chem.~Phys.}\ }\textbf {\bibinfo {volume} {130}},\ \bibinfo
  {pages} {084105} (\bibinfo {year} {2009})}\BibitemShut {NoStop}%
\bibitem [{\citenamefont {Hu}\ \emph {et~al.}(2011)\citenamefont {Hu},
  \citenamefont {Luo}, \citenamefont {Jiang}, \citenamefont {Xu},\ and\
  \citenamefont {Yan}}]{hu11}%
  \BibitemOpen
  \bibfield  {author} {\bibinfo {author} {\bibfnamefont {J.}~\bibnamefont
  {Hu}}, \bibinfo {author} {\bibfnamefont {M.}~\bibnamefont {Luo}}, \bibinfo
  {author} {\bibfnamefont {F.}~\bibnamefont {Jiang}}, \bibinfo {author}
  {\bibfnamefont {R.-X.}\ \bibnamefont {Xu}},\ and\ \bibinfo {author}
  {\bibfnamefont {Y.-J.}\ \bibnamefont {Yan}},\ }\bibfield  {title} {\bibinfo
  {title} {Pad\'e spectrum decompositions of quantum distribution functions and
  optimal hierarchical equations of motion construction for quantum open
  systems},\ }\href@noop {} {\bibfield  {journal} {\bibinfo  {journal}
  {J.~Chem.~Phys.}\ }\textbf {\bibinfo {volume} {134}},\ \bibinfo {pages}
  {244106} (\bibinfo {year} {2011})}\BibitemShut {NoStop}%
\bibitem [{\citenamefont {Erpenbeck}\ \emph {et~al.}(2018)\citenamefont
  {Erpenbeck}, \citenamefont {Hertlein}, \citenamefont {Schinabeck},\ and\
  \citenamefont {Thoss}}]{erpenbeck2018extending}%
  \BibitemOpen
  \bibfield  {author} {\bibinfo {author} {\bibfnamefont {A.}~\bibnamefont
  {Erpenbeck}}, \bibinfo {author} {\bibfnamefont {C.}~\bibnamefont {Hertlein}},
  \bibinfo {author} {\bibfnamefont {C.}~\bibnamefont {Schinabeck}},\ and\
  \bibinfo {author} {\bibfnamefont {M.}~\bibnamefont {Thoss}},\ }\bibfield
  {title} {\bibinfo {title} {Extending the hierarchical quantum master equation
  approach to low temperatures and realistic band structures},\ }\href@noop {}
  {\bibfield  {journal} {\bibinfo  {journal} {J.~Chem.~Phys.}\ }\textbf
  {\bibinfo {volume} {149}},\ \bibinfo {pages} {064106} (\bibinfo {year}
  {2018})}\BibitemShut {NoStop}%
\bibitem [{\citenamefont {Nakatsukasa}\ \emph {et~al.}(2018)\citenamefont
  {Nakatsukasa}, \citenamefont {S{\`e}te},\ and\ \citenamefont
  {Trefethen}}]{nakatsukasa2018aaa}%
  \BibitemOpen
  \bibfield  {author} {\bibinfo {author} {\bibfnamefont {Y.}~\bibnamefont
  {Nakatsukasa}}, \bibinfo {author} {\bibfnamefont {O.}~\bibnamefont
  {S{\`e}te}},\ and\ \bibinfo {author} {\bibfnamefont {L.~N.}\ \bibnamefont
  {Trefethen}},\ }\bibfield  {title} {\bibinfo {title} {The {AAA} algorithm for
  rational approximation},\ }\href@noop {} {\bibfield  {journal} {\bibinfo
  {journal} {SIAM J. Sci. Comput.}\ }\textbf {\bibinfo {volume} {40}},\
  \bibinfo {pages} {A1494} (\bibinfo {year} {2018})}\BibitemShut {NoStop}%
\bibitem [{\citenamefont {Nakatsukasa}\ and\ \citenamefont
  {Trefethen}(2020)}]{nakatsukasa2020algorithm}%
  \BibitemOpen
  \bibfield  {author} {\bibinfo {author} {\bibfnamefont {Y.}~\bibnamefont
  {Nakatsukasa}}\ and\ \bibinfo {author} {\bibfnamefont {L.~N.}\ \bibnamefont
  {Trefethen}},\ }\bibfield  {title} {\bibinfo {title} {An algorithm for real
  and complex rational minimax approximation},\ }\href@noop {} {\bibfield
  {journal} {\bibinfo  {journal} {SIAM J. Sci. Comput.}\ }\textbf {\bibinfo
  {volume} {42}},\ \bibinfo {pages} {A3157} (\bibinfo {year}
  {2020})}\BibitemShut {NoStop}%
\bibitem [{\citenamefont {Liu}\ \emph {et~al.}(2014)\citenamefont {Liu},
  \citenamefont {Zhu}, \citenamefont {Bai},\ and\ \citenamefont {Shi}}]{liu14}%
  \BibitemOpen
  \bibfield  {author} {\bibinfo {author} {\bibfnamefont {H.}~\bibnamefont
  {Liu}}, \bibinfo {author} {\bibfnamefont {L.}~\bibnamefont {Zhu}}, \bibinfo
  {author} {\bibfnamefont {S.}~\bibnamefont {Bai}},\ and\ \bibinfo {author}
  {\bibfnamefont {Q.}~\bibnamefont {Shi}},\ }\bibfield  {title} {\bibinfo
  {title} {Reduced quantum dynamics with arbitrary bath spectral densities:
  Hierarchical equations of motion based on several different bath
  decomposition schemes.},\ }\href@noop {} {\bibfield  {journal} {\bibinfo
  {journal} {J. Chem. Phys.}\ }\textbf {\bibinfo {volume} {140}},\ \bibinfo
  {pages} {134106} (\bibinfo {year} {2014})}\BibitemShut {NoStop}%
\bibitem [{\citenamefont {Tong}\ and\ \citenamefont
  {Vojta}(2006)}]{tong2006signatures}%
  \BibitemOpen
  \bibfield  {author} {\bibinfo {author} {\bibfnamefont {N.-H.}\ \bibnamefont
  {Tong}}\ and\ \bibinfo {author} {\bibfnamefont {M.}~\bibnamefont {Vojta}},\
  }\bibfield  {title} {\bibinfo {title} {Signatures of a noise-induced quantum
  phase transition in a mesoscopic metal ring},\ }\href@noop {} {\bibfield
  {journal} {\bibinfo  {journal} {Phys.~Rev.~Lett.}\ }\textbf {\bibinfo
  {volume} {97}},\ \bibinfo {pages} {016802} (\bibinfo {year}
  {2006})}\BibitemShut {NoStop}%
\bibitem [{\citenamefont {Shnirman}\ \emph {et~al.}(2002)\citenamefont
  {Shnirman}, \citenamefont {Makhlin},\ and\ \citenamefont
  {Sch{\"o}n}}]{shnirman2002}%
  \BibitemOpen
  \bibfield  {author} {\bibinfo {author} {\bibfnamefont {A.}~\bibnamefont
  {Shnirman}}, \bibinfo {author} {\bibfnamefont {Y.}~\bibnamefont {Makhlin}},\
  and\ \bibinfo {author} {\bibfnamefont {G.}~\bibnamefont {Sch{\"o}n}},\
  }\bibfield  {title} {\bibinfo {title} {Noise and decoherence in quantum
  two-level systems},\ }\href@noop {} {\bibfield  {journal} {\bibinfo
  {journal} {Phys. Scr.}\ }\textbf {\bibinfo {volume} {2002}},\ \bibinfo
  {pages} {147} (\bibinfo {year} {2002})}\BibitemShut {NoStop}%
\bibitem [{\citenamefont {Yamamoto}\ and\ \citenamefont
  {Kato}(2019)}]{yamamoto2019microwave}%
  \BibitemOpen
  \bibfield  {author} {\bibinfo {author} {\bibfnamefont {T.}~\bibnamefont
  {Yamamoto}}\ and\ \bibinfo {author} {\bibfnamefont {T.}~\bibnamefont
  {Kato}},\ }\bibfield  {title} {\bibinfo {title} {Microwave scattering in the
  subohmic spin-boson systems of superconducting circuits},\ }\href@noop {}
  {\bibfield  {journal} {\bibinfo  {journal} {J. Phys. Soc. Jpn.}\ }\textbf
  {\bibinfo {volume} {88}},\ \bibinfo {pages} {094601} (\bibinfo {year}
  {2019})}\BibitemShut {NoStop}%
\bibitem [{Note2()}]{Note2}%
  \BibitemOpen
  \bibinfo {note} {The small shift between the analytical prediction and
  FP-HEOM is mainly due to the $\alpha \to 0$-limit of the prefactor in
  Eq.(\ref {eq:shiba-subohmic})}\BibitemShut {NoStop}%
\bibitem [{\citenamefont {Wang}\ and\ \citenamefont
  {Thoss}(2010)}]{wang10from}%
  \BibitemOpen
  \bibfield  {author} {\bibinfo {author} {\bibfnamefont {H.}~\bibnamefont
  {Wang}}\ and\ \bibinfo {author} {\bibfnamefont {M.}~\bibnamefont {Thoss}},\
  }\bibfield  {title} {\bibinfo {title} {From coherent motion to localization:
  {II}. dynamics of the spin-boson model with sub-ohmic spectral density at
  zero temperature},\ }\href@noop {} {\bibfield  {journal} {\bibinfo  {journal}
  {Chem.~Phys.}\ }\textbf {\bibinfo {volume} {370}},\ \bibinfo {pages} {78}
  (\bibinfo {year} {2010})}\BibitemShut {NoStop}%
\bibitem [{\citenamefont {Ren}\ \emph {et~al.}(2022)\citenamefont {Ren},
  \citenamefont {Li}, \citenamefont {Jiang}, \citenamefont {Wang},\ and\
  \citenamefont {Shuai}}]{ren2022time}%
  \BibitemOpen
  \bibfield  {author} {\bibinfo {author} {\bibfnamefont {J.}~\bibnamefont
  {Ren}}, \bibinfo {author} {\bibfnamefont {W.}~\bibnamefont {Li}}, \bibinfo
  {author} {\bibfnamefont {T.}~\bibnamefont {Jiang}}, \bibinfo {author}
  {\bibfnamefont {Y.}~\bibnamefont {Wang}},\ and\ \bibinfo {author}
  {\bibfnamefont {Z.}~\bibnamefont {Shuai}},\ }\bibfield  {title} {\bibinfo
  {title} {Time-dependent density matrix renormalization group method for
  quantum dynamics in complex systems},\ }\href@noop {} {\bibfield  {journal}
  {\bibinfo  {journal} {Wiley Interdiscip. Rev. Comput. Mol. Sci.}\ }\textbf
  {\bibinfo {volume} {12}},\ \bibinfo {pages} {e1614} (\bibinfo {year}
  {2022})}\BibitemShut {NoStop}%
\bibitem [{\citenamefont {Kreisbeck}\ and\ \citenamefont
  {Kramer}(2012)}]{kreisbeck2012long}%
  \BibitemOpen
  \bibfield  {author} {\bibinfo {author} {\bibfnamefont {C.}~\bibnamefont
  {Kreisbeck}}\ and\ \bibinfo {author} {\bibfnamefont {T.}~\bibnamefont
  {Kramer}},\ }\bibfield  {title} {\bibinfo {title} {Long-lived electronic
  coherence in dissipative exciton dynamics of light-harvesting complexes},\
  }\href@noop {} {\bibfield  {journal} {\bibinfo  {journal} {Phys. Chem.
  Lett.}\ }\textbf {\bibinfo {volume} {3}},\ \bibinfo {pages} {2828} (\bibinfo
  {year} {2012})}\BibitemShut {NoStop}%
\bibitem [{\citenamefont {Feynman}\ and\ \citenamefont
  {Vernon}(1963)}]{feynman63}%
  \BibitemOpen
  \bibfield  {author} {\bibinfo {author} {\bibfnamefont {R.}~\bibnamefont
  {Feynman}}\ and\ \bibinfo {author} {\bibfnamefont {F.}~\bibnamefont
  {Vernon}},\ }\bibfield  {title} {\bibinfo {title} {The theory of a general
  quantum system interacting with a linear dissipative system},\ }\href
  {https://doi.org/https://doi.org/10.1016/0003-4916(63)90068-X} {\bibfield
  {journal} {\bibinfo  {journal} {Ann. Phys. (N.Y.)}\ }\textbf {\bibinfo
  {volume} {24}},\ \bibinfo {pages} {118} (\bibinfo {year} {1963})}\BibitemShut
  {NoStop}%
\bibitem [{\citenamefont {Echave}\ and\ \citenamefont
  {Clary}(1992)}]{echave92}%
  \BibitemOpen
  \bibfield  {author} {\bibinfo {author} {\bibfnamefont {J.}~\bibnamefont
  {Echave}}\ and\ \bibinfo {author} {\bibfnamefont {D.~C.}\ \bibnamefont
  {Clary}},\ }\bibfield  {title} {\bibinfo {title} {Potential optimized
  discrete variable representation},\ }\href
  {https://doi.org/https://doi.org/10.1016/0009-2614(92)85330-D} {\bibfield
  {journal} {\bibinfo  {journal} {Chem.~Phys.~Lett.}\ }\textbf {\bibinfo
  {volume} {190}},\ \bibinfo {pages} {225} (\bibinfo {year}
  {1992})}\BibitemShut {NoStop}%
\bibitem [{\citenamefont {Lubich}\ \emph {et~al.}(2015)\citenamefont {Lubich},
  \citenamefont {Oseledets},\ and\ \citenamefont {Vandereycken}}]{lubich15}%
  \BibitemOpen
  \bibfield  {author} {\bibinfo {author} {\bibfnamefont {C.}~\bibnamefont
  {Lubich}}, \bibinfo {author} {\bibfnamefont {I.}~\bibnamefont {Oseledets}},\
  and\ \bibinfo {author} {\bibfnamefont {B.}~\bibnamefont {Vandereycken}},\
  }\bibfield  {title} {\bibinfo {title} {Time integration of tensor trains},\
  }\href@noop {} {\bibfield  {journal} {\bibinfo  {journal} {SIAM J. Num.
  Anal.}\ }\textbf {\bibinfo {volume} {53}},\ \bibinfo {pages} {917} (\bibinfo
  {year} {2015})}\BibitemShut {NoStop}%
\bibitem [{\citenamefont {Haegeman}\ \emph {et~al.}(2016)\citenamefont
  {Haegeman}, \citenamefont {Lubich}, \citenamefont {Oseledets}, \citenamefont
  {Vandereycken},\ and\ \citenamefont {Verstraete}}]{haegeman2016unifying}%
  \BibitemOpen
  \bibfield  {author} {\bibinfo {author} {\bibfnamefont {J.}~\bibnamefont
  {Haegeman}}, \bibinfo {author} {\bibfnamefont {C.}~\bibnamefont {Lubich}},
  \bibinfo {author} {\bibfnamefont {I.}~\bibnamefont {Oseledets}}, \bibinfo
  {author} {\bibfnamefont {B.}~\bibnamefont {Vandereycken}},\ and\ \bibinfo
  {author} {\bibfnamefont {F.}~\bibnamefont {Verstraete}},\ }\bibfield  {title}
  {\bibinfo {title} {Unifying time evolution and optimization with matrix
  product states},\ }\href {https://doi.org/10.1103/PhysRevB.94.165116}
  {\bibfield  {journal} {\bibinfo  {journal} {Phys. Rev. B}\ }\textbf {\bibinfo
  {volume} {94}},\ \bibinfo {pages} {165116} (\bibinfo {year}
  {2016})}\BibitemShut {NoStop}%
\end{thebibliography}%


\begin{thebibliography}{10}%
\makeatletter
\providecommand \@ifxundefined [1]{%
 \@ifx{#1\undefined}
}%
\providecommand \@ifnum [1]{%
 \ifnum #1\expandafter \@firstoftwo
 \else \expandafter \@secondoftwo
 \fi
}%
\providecommand \@ifx [1]{%
 \ifx #1\expandafter \@firstoftwo
 \else \expandafter \@secondoftwo
 \fi
}%
\providecommand \natexlab [1]{#1}%
\providecommand \enquote  [1]{``#1''}%
\providecommand \bibnamefont  [1]{#1}%
\providecommand \bibfnamefont [1]{#1}%
\providecommand \citenamefont [1]{#1}%
\providecommand \href@noop [0]{\@secondoftwo}%
\providecommand \href [0]{\begingroup \@sanitize@url \@href}%
\providecommand \@href[1]{\@@startlink{#1}\@@href}%
\providecommand \@@href[1]{\endgroup#1\@@endlink}%
\providecommand \@sanitize@url [0]{\catcode `\\12\catcode `\$12\catcode
  `\&12\catcode `\#12\catcode `\^12\catcode `\_12\catcode `\%12\relax}%
\providecommand \@@startlink[1]{}%
\providecommand \@@endlink[0]{}%
\providecommand \url  [0]{\begingroup\@sanitize@url \@url }%
\providecommand \@url [1]{\endgroup\@href {#1}{\urlprefix }}%
\providecommand \urlprefix  [0]{URL }%
\providecommand \Eprint [0]{\href }%
\providecommand \doibase [0]{https://doi.org/}%
\providecommand \selectlanguage [0]{\@gobble}%
\providecommand \bibinfo  [0]{\@secondoftwo}%
\providecommand \bibfield  [0]{\@secondoftwo}%
\providecommand \translation [1]{[#1]}%
\providecommand \BibitemOpen [0]{}%
\providecommand \bibitemStop [0]{}%
\providecommand \bibitemNoStop [0]{.\EOS\space}%
\providecommand \EOS [0]{\spacefactor3000\relax}%
\providecommand \BibitemShut  [1]{\csname bibitem#1\endcsname}%
\let\auto@bib@innerbib\@empty
\bibitem [{\citenamefont {Nakatsukasa}\ \emph {et~al.}(2018)\citenamefont
  {Nakatsukasa}, \citenamefont {S{\`e}te},\ and\ \citenamefont
  {Trefethen}}]{nakatsukasa2018aaa}%
  \BibitemOpen
  \bibfield  {author} {\bibinfo {author} {\bibfnamefont {Y.}~\bibnamefont
  {Nakatsukasa}}, \bibinfo {author} {\bibfnamefont {O.}~\bibnamefont
  {S{\`e}te}},\ and\ \bibinfo {author} {\bibfnamefont {L.~N.}\ \bibnamefont
  {Trefethen}},\ }\bibfield  {title} {\bibinfo {title} {The {AAA} algorithm for
  rational approximation},\ }\href@noop {} {\bibfield  {journal} {\bibinfo
  {journal} {SIAM J. Sci. Comput.}\ }\textbf {\bibinfo {volume} {40}},\
  \bibinfo {pages} {A1494} (\bibinfo {year} {2018})}\BibitemShut {NoStop}%
\bibitem [{\citenamefont {Shi}\ \emph {et~al.}(2018)\citenamefont {Shi},
  \citenamefont {Xu}, \citenamefont {Yan},\ and\ \citenamefont
  {Xu}}]{shi2018efficient}%
  \BibitemOpen
  \bibfield  {author} {\bibinfo {author} {\bibfnamefont {Q.}~\bibnamefont
  {Shi}}, \bibinfo {author} {\bibfnamefont {Y.}~\bibnamefont {Xu}}, \bibinfo
  {author} {\bibfnamefont {Y.}~\bibnamefont {Yan}},\ and\ \bibinfo {author}
  {\bibfnamefont {M.}~\bibnamefont {Xu}},\ }\bibfield  {title} {\bibinfo
  {title} {Efficient propagation of the hierarchical equations of motion using
  the matrix product state method},\ }\href@noop {} {\bibfield  {journal}
  {\bibinfo  {journal} {J.~Chem.~Phys.}\ }\textbf {\bibinfo {volume} {148}},\
  \bibinfo {pages} {174102} (\bibinfo {year} {2018})}\BibitemShut {NoStop}%
\bibitem [{\citenamefont {Shi}\ \emph {et~al.}(2009)\citenamefont {Shi},
  \citenamefont {Chen}, \citenamefont {Nan}, \citenamefont {Xu},\ and\
  \citenamefont {Yan}}]{shi2009efficient}%
  \BibitemOpen
  \bibfield  {author} {\bibinfo {author} {\bibfnamefont {Q.}~\bibnamefont
  {Shi}}, \bibinfo {author} {\bibfnamefont {L.}~\bibnamefont {Chen}}, \bibinfo
  {author} {\bibfnamefont {G.}~\bibnamefont {Nan}}, \bibinfo {author}
  {\bibfnamefont {R.-X.}\ \bibnamefont {Xu}},\ and\ \bibinfo {author}
  {\bibfnamefont {Y.}~\bibnamefont {Yan}},\ }\bibfield  {title} {\bibinfo
  {title} {Efficient hierarchical {Liouville} space propagator to quantum
  dissipative dynamics},\ }\href@noop {} {\bibfield  {journal} {\bibinfo
  {journal} {J.~Chem.~Phys.}\ }\textbf {\bibinfo {volume} {130}},\ \bibinfo
  {pages} {084105} (\bibinfo {year} {2009})}\BibitemShut {NoStop}%
\bibitem [{\citenamefont {Feynman}\ and\ \citenamefont
  {Vernon}(1963)}]{feynman63}%
  \BibitemOpen
  \bibfield  {author} {\bibinfo {author} {\bibfnamefont {R.}~\bibnamefont
  {Feynman}}\ and\ \bibinfo {author} {\bibfnamefont {F.}~\bibnamefont
  {Vernon}},\ }\bibfield  {title} {\bibinfo {title} {The theory of a general
  quantum system interacting with a linear dissipative system},\ }\href
  {https://doi.org/https://doi.org/10.1016/0003-4916(63)90068-X} {\bibfield
  {journal} {\bibinfo  {journal} {Ann. Phys. (N.Y.)}\ }\textbf {\bibinfo
  {volume} {24}},\ \bibinfo {pages} {118} (\bibinfo {year} {1963})}\BibitemShut
  {NoStop}%
\bibitem [{\citenamefont {Echave}\ and\ \citenamefont
  {Clary}(1992)}]{echave92}%
  \BibitemOpen
  \bibfield  {author} {\bibinfo {author} {\bibfnamefont {J.}~\bibnamefont
  {Echave}}\ and\ \bibinfo {author} {\bibfnamefont {D.~C.}\ \bibnamefont
  {Clary}},\ }\bibfield  {title} {\bibinfo {title} {Potential optimized
  discrete variable representation},\ }\href
  {https://doi.org/https://doi.org/10.1016/0009-2614(92)85330-D} {\bibfield
  {journal} {\bibinfo  {journal} {Chem.~Phys.~Lett.}\ }\textbf {\bibinfo
  {volume} {190}},\ \bibinfo {pages} {225} (\bibinfo {year}
  {1992})}\BibitemShut {NoStop}%
\bibitem [{\citenamefont {Borrelli}(2019)}]{borrelli2019density}%
  \BibitemOpen
  \bibfield  {author} {\bibinfo {author} {\bibfnamefont {R.}~\bibnamefont
  {Borrelli}},\ }\bibfield  {title} {\bibinfo {title} {Density matrix dynamics
  in twin-formulation: An efficient methodology based on tensor-train
  representation of reduced equations of motion},\ }\href@noop {} {\bibfield
  {journal} {\bibinfo  {journal} {J.~Chem.~Phys.}\ }\textbf {\bibinfo {volume}
  {150}},\ \bibinfo {pages} {234102} (\bibinfo {year} {2019})}\BibitemShut
  {NoStop}%
\bibitem [{\citenamefont {Lubich}\ \emph {et~al.}(2015)\citenamefont {Lubich},
  \citenamefont {Oseledets},\ and\ \citenamefont {Vandereycken}}]{lubich15}%
  \BibitemOpen
  \bibfield  {author} {\bibinfo {author} {\bibfnamefont {C.}~\bibnamefont
  {Lubich}}, \bibinfo {author} {\bibfnamefont {I.}~\bibnamefont {Oseledets}},\
  and\ \bibinfo {author} {\bibfnamefont {B.}~\bibnamefont {Vandereycken}},\
  }\bibfield  {title} {\bibinfo {title} {Time integration of tensor trains},\
  }\href@noop {} {\bibfield  {journal} {\bibinfo  {journal} {SIAM J. Num.
  Anal.}\ }\textbf {\bibinfo {volume} {53}},\ \bibinfo {pages} {917} (\bibinfo
  {year} {2015})}\BibitemShut {NoStop}%
\bibitem [{\citenamefont {Haegeman}\ \emph {et~al.}(2016)\citenamefont
  {Haegeman}, \citenamefont {Lubich}, \citenamefont {Oseledets}, \citenamefont
  {Vandereycken},\ and\ \citenamefont {Verstraete}}]{haegeman2016unifying}%
  \BibitemOpen
  \bibfield  {author} {\bibinfo {author} {\bibfnamefont {J.}~\bibnamefont
  {Haegeman}}, \bibinfo {author} {\bibfnamefont {C.}~\bibnamefont {Lubich}},
  \bibinfo {author} {\bibfnamefont {I.}~\bibnamefont {Oseledets}}, \bibinfo
  {author} {\bibfnamefont {B.}~\bibnamefont {Vandereycken}},\ and\ \bibinfo
  {author} {\bibfnamefont {F.}~\bibnamefont {Verstraete}},\ }\bibfield  {title}
  {\bibinfo {title} {Unifying time evolution and optimization with matrix
  product states},\ }\href {https://doi.org/10.1103/PhysRevB.94.165116}
  {\bibfield  {journal} {\bibinfo  {journal} {Phys. Rev. B}\ }\textbf {\bibinfo
  {volume} {94}},\ \bibinfo {pages} {165116} (\bibinfo {year}
  {2016})}\BibitemShut {NoStop}%
\bibitem [{\citenamefont {Ren}\ \emph {et~al.}(2022)\citenamefont {Ren},
  \citenamefont {Li}, \citenamefont {Jiang}, \citenamefont {Wang},\ and\
  \citenamefont {Shuai}}]{ren2022time}%
  \BibitemOpen
  \bibfield  {author} {\bibinfo {author} {\bibfnamefont {J.}~\bibnamefont
  {Ren}}, \bibinfo {author} {\bibfnamefont {W.}~\bibnamefont {Li}}, \bibinfo
  {author} {\bibfnamefont {T.}~\bibnamefont {Jiang}}, \bibinfo {author}
  {\bibfnamefont {Y.}~\bibnamefont {Wang}},\ and\ \bibinfo {author}
  {\bibfnamefont {Z.}~\bibnamefont {Shuai}},\ }\bibfield  {title} {\bibinfo
  {title} {Time-dependent density matrix renormalization group method for
  quantum dynamics in complex systems},\ }\href@noop {} {\bibfield  {journal}
  {\bibinfo  {journal} {Wiley Interdiscip. Rev. Comput. Mol. Sci.}\ }\textbf
  {\bibinfo {volume} {12}},\ \bibinfo {pages} {e1614} (\bibinfo {year}
  {2022})}\BibitemShut {NoStop}%
\bibitem [{\citenamefont {Wang}\ and\ \citenamefont
  {Thoss}(2010)}]{wang10from}%
  \BibitemOpen
  \bibfield  {author} {\bibinfo {author} {\bibfnamefont {H.}~\bibnamefont
  {Wang}}\ and\ \bibinfo {author} {\bibfnamefont {M.}~\bibnamefont {Thoss}},\
  }\bibfield  {title} {\bibinfo {title} {From coherent motion to localization:
  {II}. dynamics of the spin-boson model with sub-ohmic spectral density at
  zero temperature},\ }\href@noop {} {\bibfield  {journal} {\bibinfo  {journal}
  {Chem.~Phys.}\ }\textbf {\bibinfo {volume} {370}},\ \bibinfo {pages} {78}
  (\bibinfo {year} {2010})}\BibitemShut {NoStop}%
\end{thebibliography}%

\end{document}


\title{Supplemental Information\\
Taming quantum noise for efficient low temperature simulations\\
of open quantum systems}

\author{Meng Xu}
\affiliation{
Institute  for Complex Quantum Systems and IQST, Ulm University - Albert-Einstein-Allee 11, D-89069  Ulm, Germany}
\author{Yaming Yan}
\affiliation{Beijing National Laboratory for Molecular Sciences, State Key Laboratory for Structural Chemistry of Unstable
and Stable Species, Institute of Chemistry, Chinese Academy of Sciences, Zhongguancun, Beijing 100190, China and
University of Chinese Academy of Sciences, Beijing 100049, China}
\author{Qiang Shi}
\affiliation{Beijing National Laboratory for Molecular Sciences, State Key Laboratory for Structural Chemistry of Unstable
and Stable Species, Institute of Chemistry, Chinese Academy of Sciences, Zhongguancun, Beijing 100190, China and
University of Chinese Academy of Sciences, Beijing 100049, China}
\author{J. Ankerhold}
\affiliation{
Institute  for Complex Quantum Systems and IQST, Ulm University - Albert-Einstein-Allee 11, D-89069  Ulm, Germany}
\author{J. T. Stockburger}
\affiliation{
Institute  for Complex Quantum Systems and IQST, Ulm University - Albert-Einstein-Allee 11, D-89069  Ulm, Germany}

\date{\today}
\maketitle

In this Supplemental Information we present more information about the optimized pole distribution to represent general reservoir power spectra $S_{\beta}(\omega)$ and the derivation of the FP-HEOM from the Feynman-Vernon path integral expression of the reduced density. We also provide more material to illustrate the efficiency, accuracy, and broad applicability of the FP-HEOM.

\section{Optimizing free poles via rational barycentric representation}
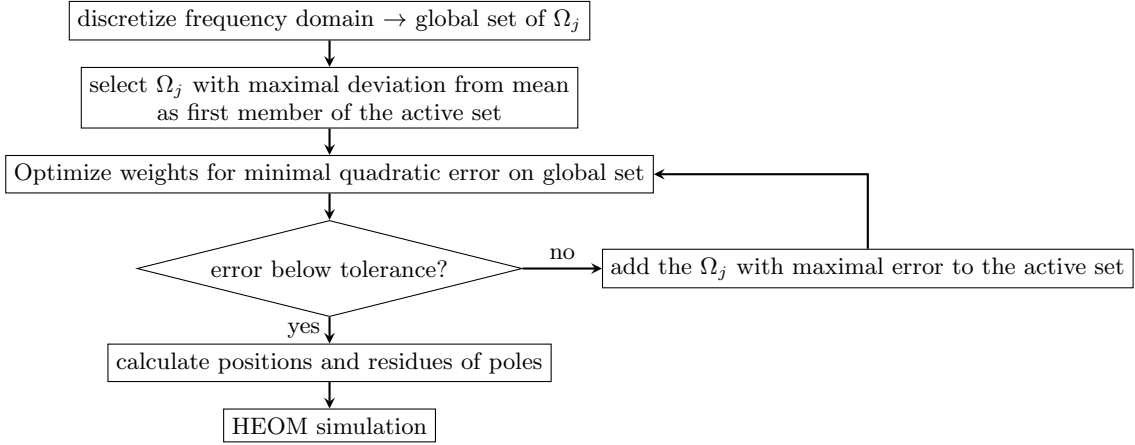
\begin{figure}[htbp]
\centering
\begin{tikzpicture}[node distance=10pt]
  \node[process]                                     (discretize)  {discretize frequency domain $\to$ global set of $\Omega_j$};
  \node[process,align=center, below=of discretize]     (init) {select $\Omega_j$ with maximal deviation from mean\\ as first member of the active set};
  \node[process, align=center, below=of init]      (weights)  {Optimize weights for minimal quadratic error on global set};
  \node[decision,below=of weights] (tolcheck)  {error below tolerance?};
  \node[process, right=of tolcheck, xshift=20pt]      (extend)  {add the $\Omega_j$ with maximal error to the active set};
  \node[process,below=of tolcheck]              (poles)  {calculate positions and residues of poles};
  \node[process,below=of poles]                     (HEOM)  {HEOM simulation};

  \draw[arrow] (discretize) -- (init);
  \draw[arrow] (init) -- (weights);
  \draw[arrow] (weights) -- (tolcheck);
  \draw[arrow] (tolcheck) -- node[anchor=south] {no} (extend);
  \draw[arrow] (tolcheck) -- node[anchor=east] {yes} (poles);
  \draw[arrow] (extend) |- (weights);
  \draw[arrow] (poles) -- (HEOM);
\end{tikzpicture}
\caption{Flowchart for the optimized pole distribution and decomposition of the correlation $C(t)$ in the FP-HEOM.}
\label{fig5}
\end{figure}

Figure \ref{fig5} illustrates the general flowchart for the FP-HEOM including the optimized pole search. It consists of three major steps:

(i) For a given spectral noise power $S_{\beta}(\omega)$ with a finite cut-off frequency $\omega_c$, one chooses a domain $\omega\in\mathcal{A}=[-N_c\,\omega_c,N_c'\,\omega_c]$, where $N_c, N_c'$ are properly chosen integers to fully cover the range of $S_\beta(\omega)$. For an exponential cut-off or a Lorentzian-type of cut-off $N_c=N_c'=20$ is sufficient.
The frequency domain $\mathcal{A}$ is then discretized to an expected resolution using a number $N_A$ of sampling points $\Omega_j$. For example, for a sub-ohmic reservoir spectrum, a convenient choice is a logarithmic discretization as also adopted in NRG, TD-DMRG, and ML-MCTDH to put emphasis on the relevant low frequency part. These points form a large \emph{global set} from which the algorithm picks support points and which it uses for error control.

(ii) The AAA -algorithm \cite{nakatsukasa2018aaa} starts with an empty \emph{active set}, as indicated in Fig.~\ref{fig5}. Two intermediate steps have the optimization objective $\tilde S_\beta(\Omega_j)-S_\beta(\Omega_j)$,
\begin{equation}\label{Eq:rationbary1}
 \tilde{S}_{\beta}^{(m-1)} (\omega) = \left. \sum_{k}^{m} 
 \frac{W_k S_{\beta}(\Omega_k)}{\omega-\Omega_k}   \middle/
 \sum_{k}^{m}\frac{W_k}{\omega - \Omega_k} \right.,
\end{equation}
where $m$ is the number of frequencies in the active set, and the optimization is over all $\omega$ contained in the global set. At each iteration step, the $\omega$ with the largest error becomes a new $\Omega_k$ in the active set. This is followed by an optimization of the complex-valued weights $W_k$, $1\leq k\leq m$, which are determined to find the minimal quadratic error $\sum_j (\tilde S_\beta(\Omega_j)-S_\beta(\Omega_j))^2$, where $j$ indexes the global set.  The iteration terminates if the requested tolerance (denoted by $\delta$), defined in the maximum norm, is satisfied.
For details we refer to \cite{nakatsukasa2018aaa}. 

(iii) Once the accuracy is reached over the domain $\mathcal{A}$ after a finite number of  $\tilde{m}$ steps, one obtains the wanted approximant $\tilde{S}_\beta(\omega)\equiv\tilde{S}_\beta(\omega)^{(\tilde{m}-1)}\equiv S_\beta(\omega)$. The poles of this rational function are determined by, in general, $\tilde{m}-1$ roots of a polynomial of order $\tilde{m}-1$ that appears in (\ref{Eq:rationbary1}) when nominator and denominator are multiplied by $\prod_{j=1}^{\tilde{m}} (\omega-\Omega_j)$ \cite{nakatsukasa2018aaa}. The correlation function $C(t)$ is for $t\geq 0$ then determined by a subset of $K\le\tilde{m}-1$ poles and corresponding residues in the lower half of the complex plane.

\begin{figure}[htbp]
\centering
\includegraphics[width=14cm]{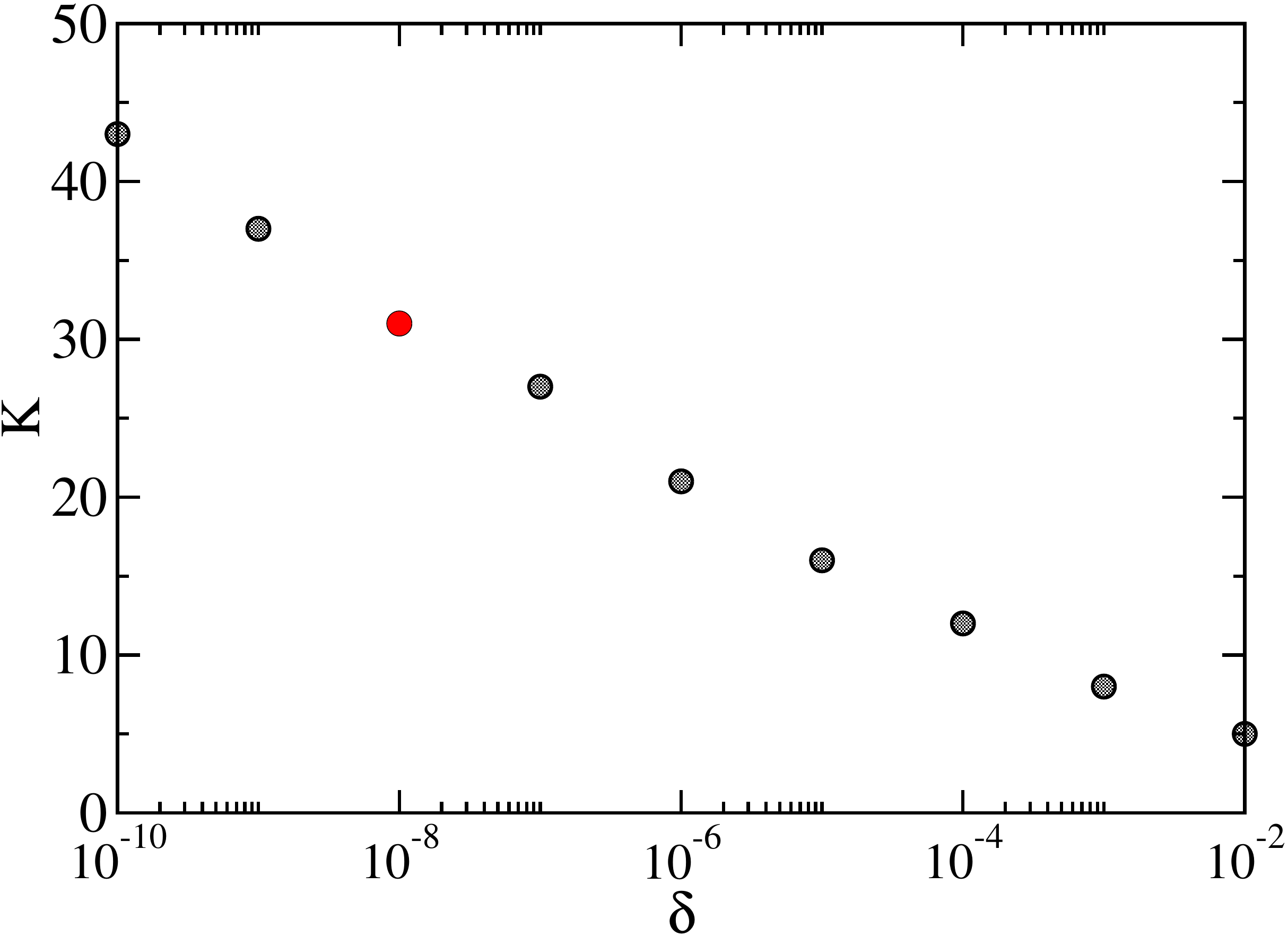}
\caption{Dependence of the number of free poles $K$ to represent the noise power on the optimization tolerance $\delta$ in sub-ohmic case. The red dot corresponds the situation used in Figs.~1 and 2 in the main text. The parameters are: $s=1/2$, $\alpha = 0.05$, $\omega_c = 20$, $\mathcal{A} = [-10^3, -10^{-4}]\cup [10^{-4},10^3]$, and $T = 0$.}
\label{fig6}
\end{figure}

When $S_{\beta}(\omega)$ is a smooth function, the discretization error is negligible for a finely spaced global set. Non-analytic behavior of $S_\beta(\omega)$ in the sub-ohmic case is addressed by the logarithmic spacing of support points near $\omega=0$. Numerical comparison verifies that the resulting multi-exponential approximation to $C(t)$ is excellent. As shown in Fig. \ref{fig6}, even with fairly tight tolerances, the number of poles at which the procedure terminates is small enough to be practical for HEOM, even at $T = 0$, with a typical scaling $K\propto |\log(\delta)|$ in the regime of very high accuracy. By using matrix product state representation the computational resources increase linearly with increasing  $K$  \cite{shi2018efficient} in contrast to the conventional HEOM \cite{shi2009efficient}. This means that FP-HEOM can provide high precision benchmarks to improve experimental set-ups on reasonable time scales, i.e. typically a few hours compared to a few days or even weeks for alternative approaches (for example, Path Integral Monte Carlo, Stochastic Liouville-von Neumann Equation).

\section{Free-Pole-HEOM}
In path integral representation \cite{feynman63} the reduced density matrix Eq. (1) in main text reads
\begin{equation}\label{Eq:pi}
  \rho(t)=\int\mathcal{D}q^{+}(t)\mathcal{D}q^{-}(t)\, {\rm e}^{i\{S_{+}[q^{+}(t)]-S_{-}[q^{-}(t)]\}}\, \mathcal{F}[q^{+}(t),q^{-}(t)]\, \rho(0) \;\; ,
\end{equation}
where $q^{+}(t)$ and $q^{-}(t)$ denote forward and backward system paths, respectively, and $S_{+}[q^+(t)]$,  $S_{-}[q^{-}(t)]$ the corresponding actions. Generally, a continuous system coordinate can be discretized using a system-specific discrete variable representation (DVR) \cite{echave92}. Specially, for the HEOM and the two level system (TLS) considered in the main text, an eigenstate representation is convenient. The effective impact of the reservoir onto the system dynamics is described by the Feynman-Vernon influence functional \cite{feynman63}, i.e.,
\begin{equation}\label{Eq:fv}
  \mathcal{F}[q^{+}(t),q^{-}(t)] 
  = \exp\left\{ -\int_{0}^{t}ds\, [q^{+}(s)-q^{-}(s)] \int_{0}^{s}d\tau
  \left[C(s-\tau)q^{+}(\tau) - C^{\ast}(s-\tau)q^{-}(\tau) \right] \right\} \;\;,
\end{equation}

Now, using the barycentric decomposition $C(t)=\sum_{k=1}^K d_k {\rm e}^{-z_k t} $ [Eq. (5) in the main text] with complex-valued coefficients $d_k$ and $z_k=i\omega_k+\gamma_k$ with real-valued frequencies $\omega_k$, $\gamma_k>0$, the path integral (\ref{Eq:pi}) can be solved with the definition of auxiliary density operators (ADOs) in forward and backward path, i.e.,
\begin{equation}\label{eq:ados2}
\begin{split}
\rho_{\bf m,n}(t)
=&\int \mathcal{D}q^+(t)\mathcal{D}q^-(t)
e^{i\{S_{+}[q^+(t)]-S_{-}[q^-(t)]\}} \prod_{k} 
\left\{-i\int_0^t d\tau\; q^{+}(\tau)\;d_k\;e^{-z_k (t-\tau)} \right\}^{m_k} \\
&\times
\left\{i\int_0^t d\tau\; q^{-}(\tau)\; d_k^{*}\;e^{-z_k^{*} (t-\tau)} \right\}^{n_k} \times
\mathcal{F}[q^+(t),q^-(t)] \; \rho(0)\;\;.
\end{split}
\end{equation}
The correspondingly scaled $\rho_{\bm{m,n}}(t)\rightarrow\prod_k\sqrt{m_k!d_k^{m_k}}\sqrt{n_k!d_k^{*n_k}}\,\rho_{\bm{m,n}}(t)$ \cite{shi2009efficient} obeys time evolution equations according to 
\begin{equation}\label{Eq:baryheom4}
\begin{split}
\frac{d}{dt} \hat{\rho}_{\bf m,n} =
&-\left(i\mathcal{L}_S
+\sum_{k=1}^{K} m_{k} z_{k} + \sum_{k=1}^{K}
n_{k} z_{k}^{*} \right) \hat{\rho}_{{\bf m,n}} 
-i\sum_{k=1}^{K}\sqrt{(m_k + 1)d_k} \left[\hat{q},\hat{\rho}_{{\bf m}_k^{+},{\bf n}} \right]  \\
&-i\sum_{k=1}^{K} \sqrt{(n_k + 1) d_k^*}\left[\hat{q},\hat{\rho}_{{\bf m,n}_k^{+}} \right]  
-i\sum_{k=1}^{K}\sqrt{m_k d_{k}}\;\hat{q}\;\hat{\rho}_{{\bf m}_k^{-},{\bf n}} 
+i\sum_{k=1}^{K}
\sqrt{n_k d_{k}^{*}}\;\hat{\rho}_{{\bf m,n}_k^{-}}\hat{q}\, .
\;\;,
\end{split}
\end{equation}
Here, the subscript $\bm{n} = \{n_1,...,n_k,...\}$ with $n_k=0, 1, 2, \ldots N_b$ indicates the definition of $\rho_{\bf n}$ and involves the $n_k$th order in $C(t)$ for the $k$th quasi-mode and $\bm{n}_k^{\pm}$ denote $\{n_1,\ldots,n_k\pm1,\ldots\}$. The multi-index $\bm{m}=\bm{n}=0$ gives the physical reduced density matrix while non-zero indices refer to ADOs. This multi-array differential equations is most conveniently propagated with a time-dependent variational principle (TDVP)\cite{shi2018efficient,borrelli2019density,lubich15,haegeman2016unifying} in matrix product states (MPS).

Upon introducing super-operators via
\begin{subequations}
\begin{equation}
    \mathcal{L}_k^+\hat{\rho}_{\bm{m,n}} \equiv \sqrt{(m_k+1)d_k}\,[\hat{q},\hat{\rho}_{\bm{m}_k^+,\bm{n}}] +                 \sqrt{(n_k+1)d_k^*}\,[\hat{q},\hat{\rho}_{\bm{m},\bm{n}_k^+}]  \;\;,
\end{equation}
\begin{equation}
    \mathcal{L}_k^-\hat{\rho}_{\bm{m,n}} \equiv \sqrt{m_kd_k}\,\hat{q}\hat{\rho}_{\bm{m}_k^-,\bm{n}} - \sqrt{n_kd_k^*}\,\hat{\rho}_{\bm{m},\bm{n}_k^-}\hat{q} \;\; ,
\end{equation}
\end{subequations}
the FP-HEOM in (\ref{Eq:baryheom4}) can be written as in the main text
\begin{equation}\label{Eq:baryheom}
   \dot{\hat{\rho}}_{\bf m,n} = -i\mathcal{L}_S\hat{\rho}_{\bf m,n} -\sum_{k=1}^K (m_k z_k+n_k z_k^*) \hat{\rho}_{\bf m,n} -i\sum_{k=1}^K \mathcal{L}_k^{+} \hat{\rho}_{\bm{m,n}} -i\sum_{k=1}^K \mathcal{L}_k^{-} \hat{\rho}_{\bm{m,n}} \;\;.
\end{equation}

We conclude that the barycentric representation of the bath-bath-correlation function allows for an  'optimized unraveling'  of the full reduced density operator with the help of auxiliarly densities in the sense that it provides a minimal basis set of non-unitary bosonic modes for an effective representation of thermal reservoirs within the HEOM. This way, one avoids the drawbacks associated with the extreme cases of either a direct evaluation of the reduced path integral, for example via Path Integral Monte Carlo (PIMC), or a basis set treatment in full (system+reservoir) Hilbert space.
While the PIMC suffers from instabilities (dynamical sign problem) due to an exponentially growing sampling space which renders long-time simulations practically not feasible, full basis set methods require a strongly growing number of modes with increasing time due to the relevance of low frequency portions of reservoirs; they thus suffer from high computational costs and truncation errors at low temperatures and stronger system-reservoir coupling. We think that the new optimized representation can also be implemented in these latter approaches to boost their performances substantially.

\section{More applications of the FP-HEOM}
We here provide additional material to the sub-ohmic spin-boson model as discussed in the main part. To highlight the efficiency of the FP-HEOM for structured reservoirs, we also provide results for relaxation of a two level system (TLS) in a bandgap environment.

\begin{figure}[htbp]
\centering
\includegraphics[width=16cm]{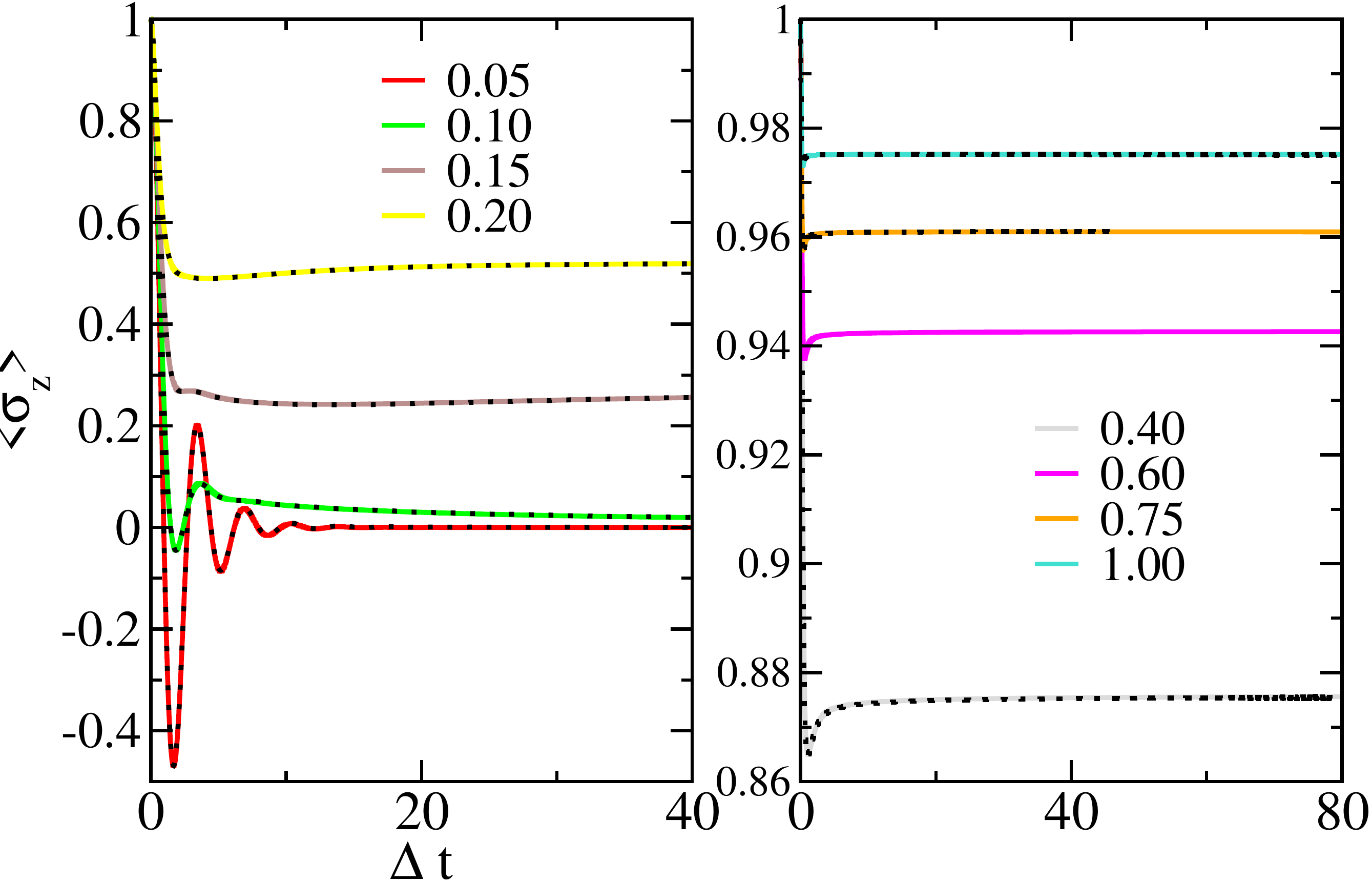}
\caption{Coherence to localization transition dynamics affected by $\alpha\in[0.05, 1.00]$. 
Solid lines denote FP-HEOM simulations, while black dot lines denote TD-DMRG simulations \cite{ren2022time}.
The simulation parameters are: $s=1/2$, $\epsilon = 0$, $\Delta = 1$, $\omega_c = 20$, and $T=0$.}
\label{fig7}
\end{figure}
(i) Figure (\ref{fig7}) displays the transition from delocalization to localization for various system-reservoir couplings $\alpha$ of the sub-ohmic spectral distribution $J(\omega)\propto \omega^{1/2}$, not shown in the main text. For TD-DMRG data were not available for all parameters.

\begin{figure}[htbp]
\centering
\includegraphics[width=12cm]{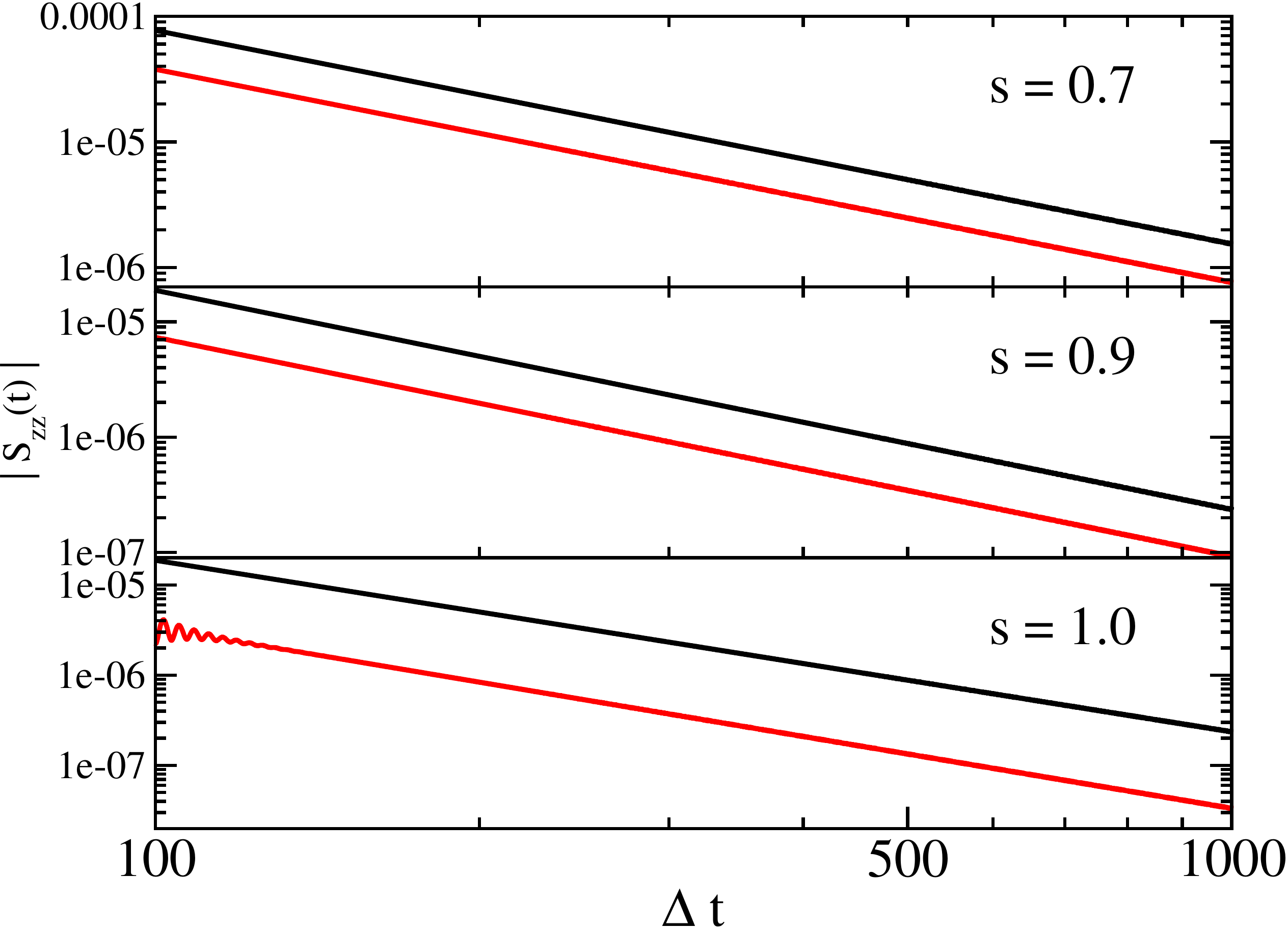}
\caption{Shiba relations in sub-Ohmic case  corresponds to $s = 0.7, 0.9, 1.0$. Black lines denote Shiba predictions, while red lines denote FP-HEOM simulations. The optimization parameters are: $\mathcal{A} = [-10^3, -10^{-4}]\cup [10^{-4},10^3]$, and accuracy $\delta = 10^{-8}$. Other simulation parameters are: $\epsilon = 0$, $\Delta = 1$, $\omega_c = 20$, $\alpha = 0.05$, ${T = 0}$.}
\label{fig8}
\end{figure}

(ii) Figure (\ref{fig8}) shows the asymptotic long-time behavior in comparison to the analytical Shiba-relation in Eq.~(10) of the main text for exponents not shown there. 
The time dependence $1/t^{1+s}$ is obtained with high precision; a small
shift between FP-HEOM data and the analytical prediction and is due to the $\alpha\to 0$-limit in its time independent prefactor.

\begin{figure}[htbp]
\centering
\includegraphics[width=12cm]{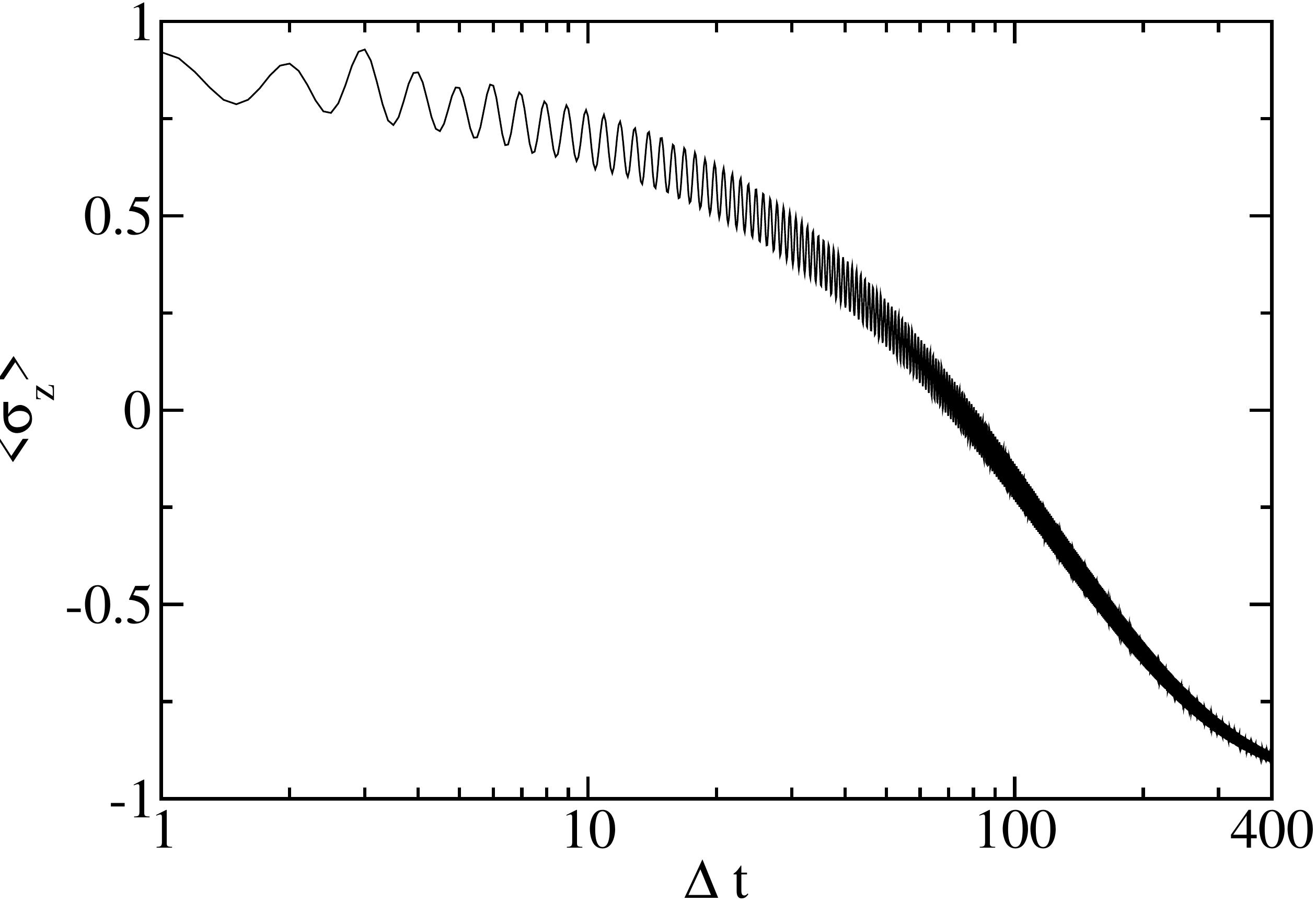}
\caption{Relaxation dynamics of a TLS in bandgap environment. Simulation parameters are: $\kappa_1 = \kappa_2 = 2$, $\xi_1 = \xi_2  = 1$, $\omega_1 = 2$, $\omega_2 = 4$, $\epsilon = 6$, and $\Delta = 1$ so that the bare TLS transition frequency is $\delta\omega=2 \sqrt{\Delta^2+\epsilon^2/4}> \omega_2>\omega_1$.}
\label{fig9}
\end{figure}

(iii) Figure (\ref{fig9}) shows the relaxation dynamics of a TLS embedded in  a bandgap environment with spectral noise power 
\begin{equation}
    S_{\beta}(\omega) 
    = \frac{\kappa_1\xi_1^8\omega}{(\omega^2-\omega_1^2)^6 + \omega^2\xi_1^{10}} + 
                        \frac{\kappa_2\xi_2^8\omega}{(\omega^2-\omega_2^2)^6 + \omega^2\xi_2^{10}} \;,\; \omega \ge 0\ 
\end{equation}
and $S_\beta(\omega)=0$ for $\omega<0$.
The parameters are chosen such that the low frequency portion of the reservoir around $\omega_1$ is well separated from the high frequency portion around $\omega_2$ and that the bare TLS transition frequency $\delta\omega=2 \sqrt{\Delta^2+\epsilon^2/4}$ sufficiently exceeds both of them. A weak coupling Lindblad treatment which reduces transition rates to be determined solely by $S_\beta(\delta\omega)$ thus predicts no impact of such a reservoir at all, while the exact FP-HEOM displays a slow relaxation towards equilibrium.

\clearpage
\pagebreak
\section{Computational costs and accuracy}

{\em (i) CPU running time:}  In order to provide more detailed information about the computational costs (CPU running time) for given parameter sets of the FP-HEOM, we collect in Table I data for the simulation results depicted in Fig.~2 and Fig.~3 of the main text. When $s = 0.5$ parameter set implemented in Fig.~2 simulations, FP-HEOM costs about 3 hours using 1 Intel Xeon Gold 6252 CPU @ 2.1 GHz core.

\begin{table}[!h]
\centering
\renewcommand{\arraystretch}{1.5}
\begin{tabular}{|m{2cm}<{\centering}|m{2.5cm}<{\centering}|m{3cm}<{\centering}|m{3cm}<{\centering}|m{2.5cm}<{\centering}|m{2cm}<{\centering}|m{2.5cm}<{\centering}|}
\hline
 $s$ & Optimization tolerance $\delta$ & FP-mode number $K$ & Bond dimension $\chi$ & Basis number $N_b$ & Simulation time\\
\hline
$s = 0.0$ & $10^{-8}$ & 31 &  40 & 10 & 5 h \\
\hline
$s=0.1$   & $10^{-8}$ & 35 &  40 & 10 & 5 h \\
\hline
$s=0.3$ & $10^{-8}$ & 39 &  40 & 10 & 5 h \\
\hline
\rowcolor{red!10} $s=0.5$ & $10^{-8}$ & 31 &  15 & 4 & 5 mins\\
\hline
$s=0.7$ & $10^{-8}$ & 19 &  30 & 10 & 1 h \\
\hline
$s=1.0$ & $10^{-8}$ & 17 &  30 & 10 & 1 h \\
\hline
\end{tabular}
\caption{Parameter details of FP-HEOM simulations in Figs.~2, 3 of the main text.}
\label{tab1}
\end{table}

\vspace{1cm}

{\em (ii) Comparison with other methods:} In Fig.~4 of the main text FP HEOM data are compared to results from TD-DMRG and ML-MCTDH data. We emphasize that the situation considered there, i.e.\ a sub-ohmic reservoir at $T=0$ for strong coupling $\alpha=0.4$,  represents an extremely challenging case for all simulation techniques of open quantum systems.  For this case neither analytical nor other numerical benchmarks are available. In Table II we provide more information about computational costs as far as they (TD-DMRG and ML-MCTDH) are available in the literature or via private communication. While a careful and thorough comparison of all three methods is beyond the scope of this paper, the data below clearly demonstrate the substantial boost in performance of the new FP-HEOM by at least one order of magnitude in CPU time.  

\begin{table}[!h]
\centering
\renewcommand{\arraystretch}{1.5}
\begin{tabular}{|c|c|c|c|}
\hline
  & FP-HEOM & TD-DMRG \cite{ren2022time} & ML-MCTDH\cite{wang10from} \\
\hline
Mode number & 16 FP-modes & 1000 Wilson’s logarithmic modes & 1000 Wilson’s logarithmic modes \\
\hline
Simulation time & 6 hours & 65 hours & up to a few days\\
\hline
Representation & MPS & MPS & Tree tensor network \\
\hline
Bond dimension & 50 & 20 & - \\
\hline 
Local site basis & 10 & 20 & 3-100 \\
\hline
Platform & \makecell[c]{1 Intel Xeon Gold\\6252 CPU @ 2.1 GHz core} &  \makecell[c]{2 Intel Xeon Gold\\ 5218R CPU @ 2.10GHz cores }& Pentium 4 @ 3 GHz \\
  \hline
\end{tabular}
\caption{Details about computational costs of the simulations in Fig.~4 in the main text.}
\label{tab2}
\end{table}

\vspace{1cm}

\begin{figure}[!h]
\centering
\includegraphics[width=16cm]{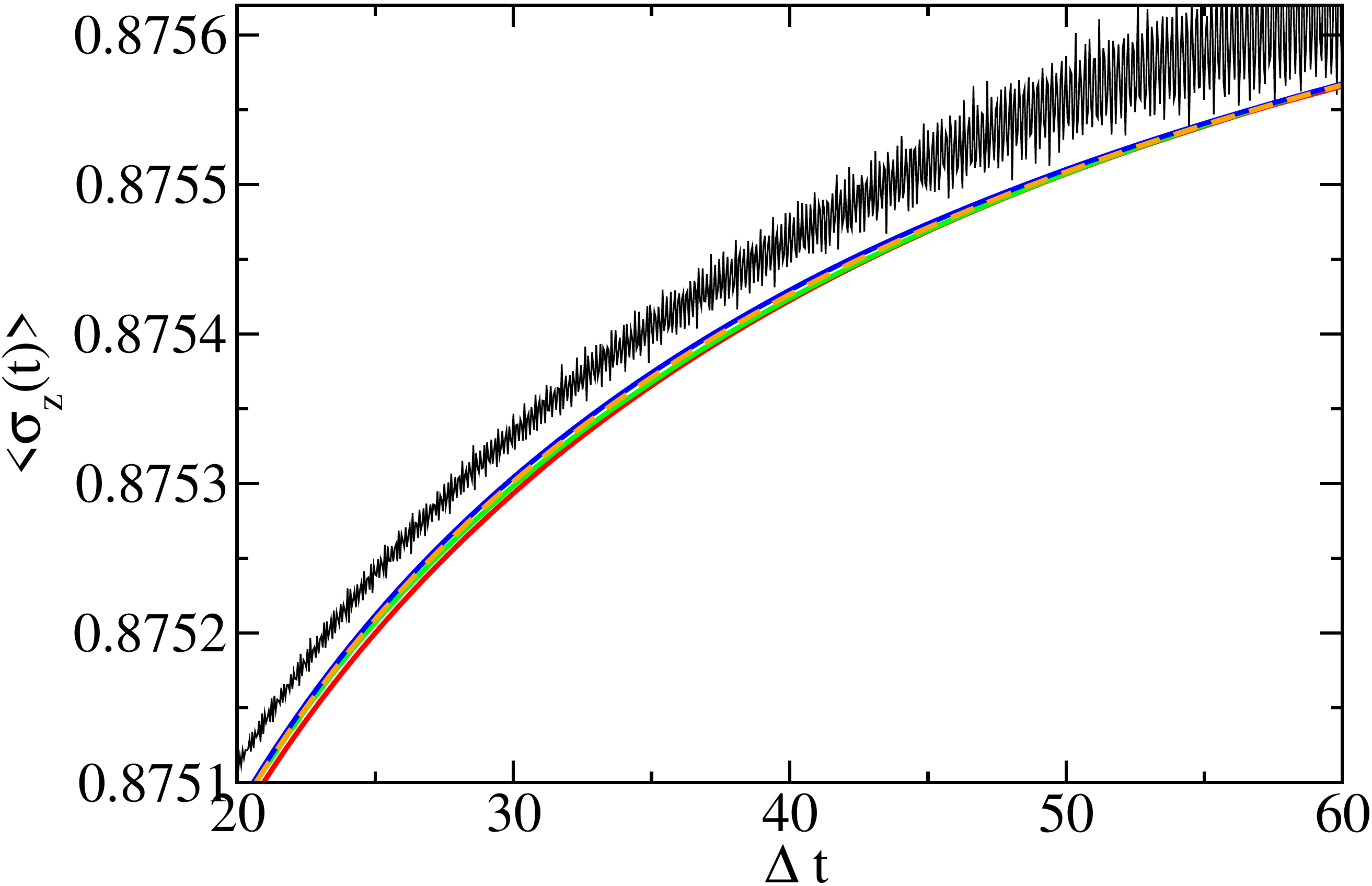}
\caption{FP-HEOM dependence on the truncation of the local site basis $N_b$. Parameter sets for the converged FP-HEOM (with accuracy $10^{-5}$) are shown in Table. \ref{tab3}. TD-DMRG  data \cite{ren2022time} are depicted in black.}
\label{fig10}
\end{figure}

\begin{figure}[!h]
\centering
\includegraphics[width=16cm]{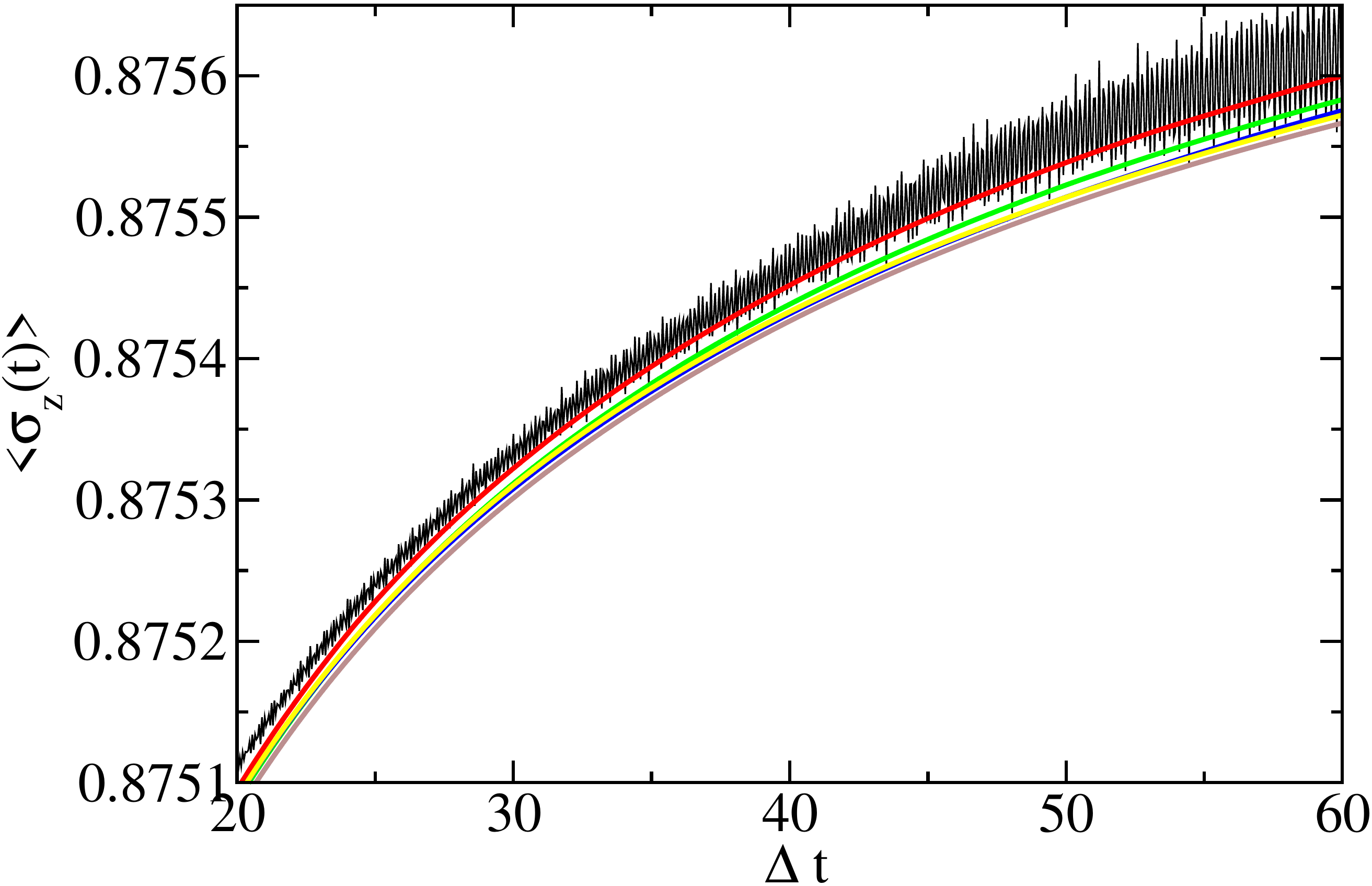}
\caption{FP-HEOM dependence on the MPS bond dimension $\chi$; parameter sets for the converged FP-HEOM (accuracy $10^{-5}$) dynamics are shown in Table. \ref{tab4}. TD-DMRG  data \cite{ren2022time} are depicted in black.}
\label{fig11}
\end{figure}

\begin{figure}[!h]
\centering
\includegraphics[width=16cm]{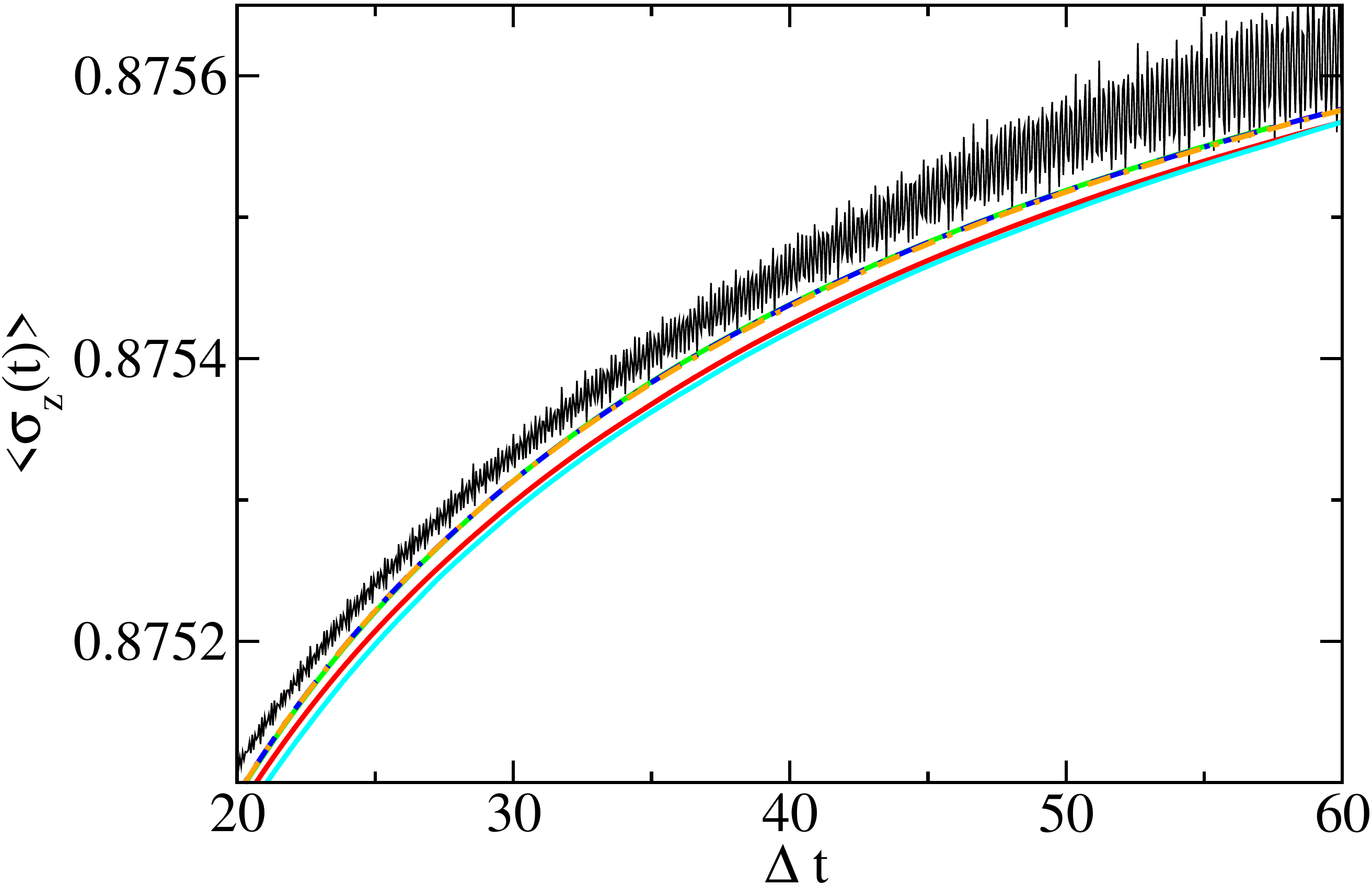}
\caption{FP-HEOM dynamics dependence on optimization tolerance $\delta$ (barycentric mode number $K$); parameter sets for the converged FP-HEOM (accuracy $10^{-5}$) dynamics are shown in Table. \ref{tab5}. Results in cyan color are shown in Fig.~4 of the main text. TD-DMRG  data \cite{ren2022time} are depicted in black.}
\label{fig12}
\end{figure}

{\em (iii) Convergence properties of FP-HEOM:} To further elucidate the convergence properties of the FP-HEOM for the situation in Fig.~4 of the main text, we show in Figs.~\ref{fig10}--\ref{fig12} and Table III--V data for various parameter sets. A few remarks are in order here: 
(a) The FP-HEOM provides converged data mainly by increasing MPS bond dimension (see Fig.~\ref{fig11}), this is well-known due to entanglement truncation.
(b) The converged FP-HEOM (up to $10^{-5}$) requires a smaller basis number $N_b$ (in MPS related to the depth of the hierarchy) even for strong system-bath coupling (see Fig.~\ref{fig10}). 
(c) Reducing the error threshold of the rational approximation to the reservoir fluctuation spectrum by three orders of magnitude ($\delta=10^{-5}\to 10^{-8}$) in combination with an increased number of FP-modes (see Fig.~\ref{fig12}) results only in a modest increase in simulation time considering a short time step $dt$ is used.
\begin{table}[!h]
\centering
\renewcommand{\arraystretch}{1.5}
\begin{tabular}{|m{1cm}<{\centering}|m{2.5cm}<{\centering}|m{3cm}<{\centering}|m{3cm}<{\centering}|m{2.5cm}<{\centering}|c |m{2.5cm}<{\centering}|}
\hline
 Color & Optimization tolerance $\delta$ & FP-mode number $K$ & Bond dimension $\chi$ & Basis number $N_b$ & Time step $dt$ & Simulation time\\
\hline
 Red & $10^{-5}$ & 16 &  60 & \cellcolor{red!20} 10 & 0.01 & 9 h \\
 \hline
 green & $10^{-5}$ & 16 &  60 & \cellcolor{red!20} 13 & 0.01 & 9 h \\
 \hline
 Blue & $10^{-5}$ & 16 &  60 & \cellcolor{red!20} 15 & 0.01 & 13 h \\
 \hline
 Orange & $10^{-5}$ & 16 &  60 & \cellcolor{red!20} 17 & 0.01 & 16 h \\
\hline
\end{tabular}
\caption{Convergence of FP-HEOM  for various basis numbers $N_b$ in Fig.~\ref{fig10}.}
\label{tab3}
\end{table}

\vspace{1cm}

\begin{table}[!h]
\centering
\renewcommand{\arraystretch}{1.5}
\begin{tabular}{|m{1cm}<{\centering}|m{2.5cm}<{\centering}|m{3cm}<{\centering}|m{3cm}<{\centering}|m{2.5cm}<{\centering}|c |m{2.5cm}<{\centering}|}
\hline
Color & Optimization tolerance $\delta$ & FP-mode number $K$ & Bond dimension $\chi$ & Basis number $N_b$ & Time step $dt$ & Simulation time\\
 \hline
Red & $10^{-5}$ & 16 & \cellcolor{blue!20} 40 & 17 & 0.01 & 6 h \\
  \hline
Green & $10^{-5}$ & 16 & \cellcolor{blue!20} 45 & 17 & 0.01 & 8 h \\
  \hline
Blue & $10^{-5}$ & 16 & \cellcolor{blue!20} 50 & 17 & 0.01 & 11 h \\
  \hline
Yellow & $10^{-5}$ & 16 & \cellcolor{blue!20} 55 & 17 & 0.01 & 12 h \\
  \hline
Brown & $10^{-5}$ & 16 & \cellcolor{blue!20} 60 & 17 & 0.01 & 16 h \\
\hline
\end{tabular}
\caption{Convergence of FP-HEOM for various bond dimensions of the MPS as shown in Fig.~\ref{fig11}.}
\label{tab4}
\end{table}

\vspace{1cm}

\begin{table}[!h]
\centering
\renewcommand{\arraystretch}{1.5}
\begin{tabular}{|m{1cm}<{\centering}|m{2.5cm}<{\centering}|m{3cm}<{\centering}|m{3cm}<{\centering}|m{2.5cm}<{\centering}|c |m{2.5cm}<{\centering}|}
\hline
Color & Optimization tolerance $\delta$ & FP-mode number $K$ & Bond dimension $\chi$ & Basis number $N_b$ & Time step $dt$ & Simulation time\\
\hline
\rowcolor{cyan!50} Cyan & $10^{-5}$ & 16 &50 & 10 & 0.01 & 6 h \\
\hline
Red & \cellcolor{green!20} $10^{-5}$ & 16 &  60 & 13 & 0.01 & 9 h \\
 \hline
Green & \cellcolor{green!20} $10^{-6}$ & 21 &  60 & 13 & 0.01 & 16 h \\
 \hline
Blue & \cellcolor{green!20} $10^{-7}$ & 27 &  60 & 13 & 0.01 & 20 h \\
 \hline
Orange &\cellcolor{green!20}  $10^{-8}$ & 31 &  60 & 13 & 0.005\footnote{Time step convergence required by the high frequency FP-modes in strong system-bath coupling.} & 38 h \\
\hline
\end{tabular}
\caption{Convergence of FP-HEOM for increasing optimization tolerance $\delta$ of the barycentric representation which directly determines the mode number $K$ as shown in Fig.~\ref{fig12}.}
\label{tab5}
\end{table}

\clearpage
\pagebreak
\section{Pole distribution in Fig.~1 of the main text}
We here collect in Table~\ref{table:quasi} the data for the optimized poles ( $\xi$ denotes poles, while $\eta$ denotes residues), which approximate $S_{\beta}(\omega)$ through
\begin{equation}
    S_{\beta}(\omega) \approx \sum_{j=1}^{31} \frac{\eta_j}{\omega - \xi_j} + c.c. \;\;.
\end{equation}
\begin{table}[!ht]
    \centering
    \begin{tabular}{|c||c|c|c|c|}
    \hline
    Number & $\Re(\eta)$ & $\Im(\eta)$ & $\Re(\xi)$ & $\Im(\xi)$ \\
    \hline
     1&   -0.007050248 & 0.002482056 & 188.7752364 & -164.9169644 \\ \hline
     2&   0.102762949 & 0.116652539 & 114.5230822 & -130.046728 \\ \hline
     3&   0.581303135 & -0.644477046 & 69.37948966 & -100.2912245 \\ \hline
     4&   -1.455575575 & -1.650266045 & 40.55914002 & -74.97480723 \\ \hline
     5&   -2.988444895 & 1.187910091 & 22.52508418 & -54.17566836 \\ \hline
     6&   -0.622081237 & 3.115931332 & 11.81278425 & -37.82551602 \\ \hline
     7&   1.30388447 & 2.013121889 & 5.852688004 & -25.57526889 \\ \hline
     8&   1.382279145 & 0.625686611 & 2.752595646 & -16.83290689 \\ \hline
     9&   0.8611309 & 0.016022 & 1.241895418 & -10.8518732 \\ \hline
    10&   0.443366368 & -0.118339778 & 0.548086947 & -6.885813753 \\ \hline
    11&    0.212002544 & -0.101557778 & 0.243348806 & -4.31568345 \\ \hline
    12&    0.09893136 & -0.063028266 & 0.113960488 & -2.680510104 \\ \hline
    13&   0.046158834 & -0.034415289 & 0.059450913 & -1.654822486 \\ \hline
    14&    0.021773639 & -0.017637193 & 0.033910866 & -1.017022414 \\ \hline
    15&    0.010348793 & -0.008760985 & 0.019524006 & -0.62200475 \\ \hline
    16&    0.004904562 & -0.004276533 & 0.010782038 & -0.378232958 \\ \hline
    17&    0.002297321 & -0.002058266 & 0.005782001 & -0.228876357 \\ \hline
    18&    0.001065577 & -0.00097944 & 0.003293429 & -0.138297459 \\ \hline
    19&    0.00049681 & -0.000464088 & 0.002202629 & -0.083659132 \\ \hline
    20&    0.000235641 & -0.000219394 & 0.001675626 & -0.050533949 \\ \hline
    21&    0.000112977 & -0.000102543 & 0.001268055 & -0.03029634 \\ \hline
    22&    5.37E-05 & -4.70E-05 & 0.000861682 & -0.017943686 \\ \hline
    23&    2.49E-05 & -2.12E-05 & 0.000496313 & -0.010482398 \\ \hline
    24&    1.12E-05 & -9.49E-06 & 0.000227853 & -0.006041449 \\ \hline
    25&    4.85E-06 & -4.23E-06 & 6.86E-05 & -0.003437852 \\ \hline
    26&    2.05E-06 & -1.89E-06 & -9.00E-06 & -0.001928643 \\ \hline
    27&    8.36E-07 & -8.31E-07 & -3.82E-05 & -0.001060056 \\ \hline
    28&    3.24E-07 & -3.56E-07 & -4.17E-05 & -0.000568788 \\ \hline
    29&    1.21E-07 & -1.49E-07 & -3.41E-05 & -0.000297603 \\ \hline
    30&    4.48E-08 & -6.28E-08 & -2.43E-05 & -0.000147216 \\ \hline
    31&    1.62E-08 & -2.71E-08 & -1.54E-05 & -5.76E-05 \\ \hline
    \end{tabular}
    \caption{Poles in the lower complex half plane in Fig. 1 in main text to represent the noise power.}
    \label{table:quasi}
\end{table}

\section{Data availability}
The data that support the figures within this article are available from the corresponding author upon reasonable request.

\section*{References}
\bibliography{FPHEOM}